\documentclass[aps,twocolumn,superscriptaddress,nofootinbib]{revtex4}
\usepackage{amsmath}
\usepackage{amsthm}
\usepackage{physics, mathrsfs}
\usepackage{amsfonts}
\usepackage{amssymb}
\usepackage{graphicx}   
\usepackage{float}
\usepackage[english]{babel} 
\usepackage[bookmarks=false]{hyperref}
\hypersetup{colorlinks, 
           citecolor=black,
           filecolor=black,
           linkcolor=red,
           urlcolor=blue,
           bookmarksopen=true,
           }
\usepackage{enumerate}
\newtheorem{lemma}{Lemma}
\newtheorem{theorem}{Theorem}
\newtheorem{corollary}{Corollary}
\newtheorem{proposition}{Proposition}
\newtheorem{definition}{Definition}
\newtheorem{assumption}{Assumption}
\newtheorem{problem}{Problem}
\newtheorem{algorithm}{Algorithm}
\begin{document}
\title{Segmentation-Based Regression for Quantum Neural Networks}

\author{James C. Hateley}

\begin{abstract}
Recent advances in quantum hardware motivate the development of algorithmic frameworks that integrate quantum sampling with classical inference. This work introduces a segmentation-based regression method tailored to quantum neural networks (QNNs), where real-valued outputs are encoded as base-bb digit sequences and inferred through greedy digitwise optimization. By casting the regression task as a constrained combinatorial problem over a structured digit lattice, the method replaces continuous inference with interpretable and tractable updates. A hybrid quantum–classical architecture is employed: quantum circuits generate candidate digits through projective measurement, while classical forward models evaluate these candidates based on task-specific error functionals. We formalize the algorithm from first principles, derive convergence and complexity bounds, and demonstrate its effectiveness on inverse problems involving PDE-constrained models. The resulting framework provides a robust, high-precision interface between quantum outputs and continuous scientific inference.
\end{abstract}

\maketitle
\section{Introduction}
Quantum computing has made significant strides in recent years, especially with the rise of noisy intermediate-scale quantum (NISQ) devices and the growing use of variational quantum algorithms~\cite{arute2019quantum,cerezo2021variational}. These advances have opened up new possibilities for scientific computing and machine learning. Among the most promising developments are quantum neural networks (QNNs), which serve as flexible models capable of capturing high-dimensional nonlinear relationships. They have found use in a wide range of applications, from physics-informed modeling to solving inverse problems using quantum-enhanced inference~\cite{farhi2018classification,schuld2020circuit,abbas2021power}.

Unlike their classical counterparts, QNNs produce outputs by quantum measurement, which leads to inherently discrete and stochastic results~\cite{peres1995quantum,schuld2019feature}. This poses a core challenge for regression tasks, which rely on recovering continuous parameters accurately from noisy or indirect observations. The issue becomes especially difficult in scientific problems governed by partial differential equations (PDEs), where small errors in parameters can cause large deviations in predicted outcomes~\cite{kaipio2005inverse}.

To tackle this problem, we propose a segmentation-based regression framework that recasts continuous inference as a combinatorial optimization over structured digit lattices. Instead of working with continuous variables directly, the method encodes targets as base-$b$ digit expansions and builds solutions one digit at a time. Each digit is selected to minimize a forward error functional, resulting in a flexible and interpretable way to refine precision hierarchically.

This approach fits naturally into a hybrid quantum-classical architecture. Quantum circuits are used to generate digit candidates via repeated measurement, while classical forward models—such as PDE solvers or surrogate simulations—evaluate each candidate to guide digit selection~\cite{cerezo2021variational,mitarai2018parameter}. This division of responsibilities plays to the strengths of each side: quantum hardware excels at stochastic sampling, while classical computation handles model evaluation and error analysis reliably.

In what follows, we develop the segmentation algorithm from the ground up, providing formal guarantees on convergence and runtime complexity. We also demonstrate its practical effectiveness in inverse problems involving PDE-constrained models. Overall, segmentation-based regression offers a scalable and precise method for hybrid inference, bridging quantum outputs with the demands of real-world scientific computation.

\section{Preliminaries}
\label{sec:preliminaries}
This section formalizes the foundational structures of segmentation-based regression. The method transforms the regression target $\theta \in \mathbb{R}^M$ into a discrete optimization problem over a structured digit lattice, enabling tractable inference via localized updates. By expressing real-valued parameters as base-$b$ digit expansions, the continuous regression space is replaced by a finite, resolution-tunable subset of $\mathbb{R}^M$, which is particularly well-aligned with the discrete nature of quantum measurement~\cite{schuld2020circuit, mitarai2018parameter, benedetti2019parameterized, farhi2018classification}.

We begin with a formal description of base-$b$ expansions and digit lattices. We then explain how quantum neural networks naturally generate digit-level outputs suitable for this representation. Finally, we introduce the classical forward model and associated loss functional that guide digit refinement. These components form the backbone of the hybrid inference architecture developed in later sections~\cite{cerezo2021variational, arute2019quantum, brassard2002quantum, schuld2019feature}.
\subsection{Digit Representation of Real Numbers}
Segmentation-based regression encodes real-valued parameters as finite base-$b$ expansions, inducing a hierarchical discretization of $\mathbb{R}^M$. This enables interpretable, resolution-tunable representations and structured digitwise optimization.

\begin{definition}[Base-$b$ Digit Expansion]\label{def:digit_representation}
Let $b \in \mathbb{Z}_{\geq 2}$ be a fixed digit base, and $(n,m) \in \mathbb{N}^2$ the digit resolution, with $n$ digits left of the radix point and $m$ to the right. A real number $y \in \mathbb{R}$ admits a base-$b$ expansion if
\begin{equation}
y = \sum_{i=-m}^{n} y_i b^i, \quad y_i \in \{0,1,\ldots,b-1\}.
\end{equation}
\end{definition}

\noindent This defines a finite set of representable values:

\begin{definition}[Segmented Output Space]
The scalar digit lattice is
\begin{equation}
\mathcal{Y}_{b,n,m} := \left\{ \sum_{i=-m}^{n} y_i b^i \;\middle|\; y_i \in \{0, \ldots, b-1\} \right\} \subset \mathbb{R}.
\end{equation}
For $M$-dimensional vectors, define the segmented space
\begin{equation}
\mathcal{Y}^{M}_{b,n,m} := \left\{ y = (y_1, \ldots, y_M) \in \mathbb{R}^M \;\middle|\; y_k \in \mathcal{Y}_{b,n,m} \right\}.
\end{equation}
Each component admits the expansion
\begin{equation}
y_k = \sum_{i=-m}^{n} y_{k,i} b^i, \quad y_{k,i} \in \{0,\ldots,b-1\}.
\end{equation}
\end{definition}

\noindent The total number of digits per component is the digit depth, $d := n + m + 1$, which also determines the number of update steps in the segmentation algorithm. This structure supports coarse-to-fine refinement and interpretable control over resolution.

To generalize this representation, we allow for negative digits and heterogeneous bases.

\begin{definition}[Signed Digit Expansion]\label{def:signed_digits}
In a signed expansion, each digit $y_i$ belongs to the symmetric set $\{-\lfloor b/2 \rfloor, \ldots, \lfloor (b-1)/2 \rfloor\}$. The representable range becomes symmetric about zero:
\begin{equation}
\theta_{\min}^{\text{signed}} = -\sum_{i=-m}^{n} \left\lfloor \frac{b}{2} \right\rfloor b^i, \quad
\theta_{\max}^{\text{signed}} = \sum_{i=-m}^{n} \left\lfloor \frac{b-1}{2} \right\rfloor b^i.
\end{equation}
\end{definition}

\begin{definition}[Mixed-Base Digit Lattice]\label{Def:mix_lat_base}
Let $b_{k,i} \in \mathbb{Z}_{\geq 2}$ denote the base assigned to digit position $i$ of component $k$. Then the mixed-base digit space is
\begin{multline}
\mathcal{Y}^{\text{mix}} := \Big\{ y \in \mathbb{R}^M \;\Big|\; y_k = \sum_{i=-m}^{n} y_{k,i} b_{k,i}^i,\; \\
 y_{k,i} \in \{0, \ldots, b_{k,i} - 1\} \Big\}.
\end{multline}
The total number of configurations in this space is
\begin{equation}
|\mathcal{Y}^{\text{mix}}| = \prod_{k=1}^{M} \prod_{i=-m}^{n} b_{k,i}.
\end{equation}
\end{definition}
This mixed-base formalism supports flexible, heterogeneous precision across components and digit layers, enabling adaptive trade-offs between expressivity and representation cost.
\subsection{Quantum Neural Network Output Model}

Quantum neural networks (QNNs) produce discrete outputs through projective measurement on entangled quantum registers~\cite{benedetti2019parameterized,schuld2020circuit}. These outputs naturally align with the segmented digit spaces introduced above, enabling digitwise inference based on quantum sampling statistics.

\begin{definition}[QNN Output Distribution]
Let $N : \mathbb{R}^L \to \mathcal{Y}^M_{b,n,m}$ be a trained QNN. For input data $x \in \mathbb{R}^L$, the network outputs a random vector
\begin{equation}
y \sim P_x, \quad \text{with support } \text{supp}(P_x) \subseteq \mathcal{Y}^M_{b,n,m},
\end{equation}
where $P_x$ is the probability distribution over the digit lattice induced by quantum measurement.
\end{definition}

Each output digit $y_{k,i}$ is sampled from a dedicated quantum register $R_{k,i}$, typically consisting of $\lceil \log_2 b \rceil$ qubits. Upon measurement in the computational basis, $R_{k,i}$ yields a value $j \in \{0, \ldots, b-1\}$ according to the Born rule~\cite{peres1995quantum}.

To stabilize digit inference, we perform repeated measurements and aggregate statistics to construct empirical distributions over digit values:

\begin{definition}[Empirical Digit Distribution]
Let $R_{k,i}$ be the quantum register for digit $y_{k,i}$. For $R$ repeated measurements, define the empirical frequency
\begin{equation}
f^{(j)}_{k,i} := \frac{1}{R} \sum_{r=1}^{R} \mathbb{I}[\text{measurement} = j], \quad j \in \{0, \ldots, b-1\},
\end{equation}
where $\mathbb{I}[\cdot]$ is the indicator function.
\end{definition}

\noindent The empirical distribution $\{f^{(j)}_{k,i}\}_{j=0}^{b-1}$ approximates the true quantum probabilities $\{p^{(j)}_{k,i}\}$ as $R \to \infty$ under independent trials.

These empirical distributions serve as input to the segmentation algorithm by defining candidate digit sets:
\begin{equation}
C_{k,i} \subseteq \{0, \ldots, b - 1\}
\end{equation}
\noindent $\text{typically via top-$r$ selection or thresholding on } f^{(j)}_{k,i}.$ Each register $R_{k,i}$ thus provides a localized probabilistic encoding of digit $y_{k,i}$. The digitwise architecture modularizes the quantum output, enabling scalable inference: quantum sampling proposes digit candidates, and classical evaluation selects among them. This structure supports robustness to measurement noise, entanglement, and decoherence, as digit updates are resolved independently in the classical loop~\cite{schuld2020circuit,mitarai2018parameter}.

\subsection{Forward Model and Error Functional}

To evaluate candidate digit configurations during segmentation, we introduce a classical forward model that maps parameter vectors to observable outputs. This forward map defines the error functional minimized during digit selection.

\begin{assumption}[Forward Model]
There exists a continuous and computable map
\begin{equation}
F : \mathbb{R}^M \to \mathbb{R}^L,
\end{equation}
which maps each parameter vector $y \in \mathbb{R}^M$ to a predicted observation $z = F(y) \in \mathbb{R}^L$.
\end{assumption}

\noindent The map $F$ is treated as a black-box simulator, requiring only function evaluation (not gradients), making the framework compatible with legacy solvers, PDE engines, and experimental data pipelines.

\begin{definition}[Weighted Error Functional]
Let $x \in \mathbb{R}^L$ be the observed data. For a symmetric positive semi-definite matrix $W \in \mathbb{R}^{L \times L}$, define
\begin{equation}
E(y) := \|F(y) - x\|_W^2 := (F(y) - x)^\top W (F(y) - x).
\end{equation}
\end{definition}

\noindent When $W = I$, this reduces to the standard squared Euclidean norm $\|F(y) - x\|^2$. If $W$ encodes sensor covariance, $E(y)$ becomes a Mahalanobis norm, adapting the geometry of the error space to reflect noise structure or physical units~\cite{kaipio2005inverse}.

To stabilize inference and promote interpretability, we augment the objective with regularization:

\begin{definition}[Regularized Loss Functional]
Let $R(y)$ be a task-specific regularization term, and $\lambda \geq 0$ a tuning parameter. Define the total objective as
\begin{equation}
L(y) := E(y) + \lambda R(y).
\end{equation}
\end{definition}

\noindent Regularizers $R(y)$ are typically designed to penalize excessive digit complexity or to promote sparsity, entropy minimization, or structural priors across digit positions.

\paragraph[Example: Digitwise $\ell_1$ Regularization]
A natural choice is a weighted $\ell_1$ penalty:
\begin{equation}
R(y) := \sum_{k=1}^M \sum_{i=-m}^{n} \omega_{k,i} |y_{k,i}|, \quad \omega_{k,i} \geq 0,
\end{equation}
where high weights on low-significance digits discourage overfitting and reduce unnecessary precision.

\begin{definition}[Digitwise-Local Injectivity]
Let $F : \mathbb{R}^M \to \mathbb{R}^L$ be a forward model. We say $F$ is digitwise-locally injective on $\mathcal{Y}^M_{b,n,m}$ if, for every $y \in \mathcal{Y}^M_{b,n,m}$ and every one-digit perturbation $y' = \delta^{(j)}_{k,i}(y) \neq y$, we have
\begin{equation}
F(y') \neq F(y).
\end{equation}
\end{definition}

\noindent This condition ensures that each digit update induces a measurable change in the forward output, guaranteeing local progress in the segmentation procedure.

\begin{problem}[Segmentation Regression]
Given observation $x \in \mathbb{R}^L$, forward model $F$, digit base $b$, and resolution $(n,m)$, find
\begin{equation}
\tilde{y} := \arg\min_{y \in \mathcal{Y}^M_{b,n,m}} E(y),
\end{equation}
i.e., the digitized parameter vector minimizing prediction error.
\end{problem}

This formulation converts the continuous regression task into a discrete optimization problem over a structured digit lattice. The segmentation--elimination algorithm, introduced in the next section, provides an efficient strategy for solving this problem using localized digitwise descent.

\section{Segmentation Algorithm for Regression}
\label{sec:algorithm}

We now introduce the segmentation–elimination algorithm: a structured digitwise optimization method for solving regression problems over the discrete digit lattice $\mathcal{Y}_{b,n,m}^M$. Rather than minimizing over continuous variables, the method incrementally constructs a solution by optimizing each digit in isolation, exploiting the separable structure of base-$b$ expansions. This allows for scalable, interpretable descent across the digit hierarchy, from coarse structural components to fine-resolution adjustments~\cite{ge2015escaping, mitarai2018parameter}.

Each component $y_k$ of the output vector is expressed in base-$b$ as
\begin{equation}
y_k = \sum_{i=-m}^{n} y_{k,i} b^i, \quad y_{k,i} \in \{0, \dots, b - 1\}.
\end{equation}
The algorithm proceeds by sweeping through digit positions in a fixed order--typically from most to least significant—refining the output via local updates guided by a forward error functional. This coarse-to-fine refinement schedule exploits the hierarchical contribution of digit positions, with early updates shaping the global parameter structure and later updates adding precision.
\subsection{Digitwise Optimization Framework}
At each digit position $(k,i)$, the algorithm considers a localized set of candidate substitutions $j \in C_{k,i}$ and selects the value minimizing the regularized loss. This modular architecture respects the additive nature of digit expansions and enables independent evaluation of candidate digits without gradient information or global recomputation~\cite{cerezo2021variational, schuld2019feature}.

\begin{definition}[Digit Neighborhood and Perturbation]
Fix $(k, i)$ with $k \in \{1, \dots, M\}$ and $i \in \{-m, \dots, n\}$. Let $C_{k,i} \subseteq \{0, \dots, b - 1\}$ be the set of candidate digits at position $(k, i)$. For each $j \in C_{k,i}$, define a perturbation map
\[
\delta_{k,i}^{(j)} : \mathcal{Y}_{b,n,m}^M \to \mathcal{Y}_{b,n,m}^M
\]
that replaces $y_{k,i}$ with $j$ while keeping all other digits fixed. The locally modified output is denoted $y_{k,i}^{(j)} := \delta_{k,i}^{(j)}(y)$.
\end{definition}

Each perturbation $\delta^{(j)}_{k,i}(y)$ defines a local neighbor of $y$ obtained by replacing a single digit $y_{k,i}$ with candidate value $j \in C_{k,i}$, while keeping all other digits fixed. The set $\{\delta^{(j)}_{k,i}(y) \mid j \in C_{k,i}\}$ thus forms the digitwise neighborhood explored during the update at position $(k,i)$.

The update rule for digit $(k,i)$ is given by
\begin{equation}
y_{k,i} \leftarrow \arg\min_{j \in C_{k,i}} \mathcal{L}(\delta_{k,i}^{(j)}(y)),
\end{equation}
where $\mathcal{L}(y) = E(y) + \lambda R(y)$ is the regularized loss functional. This rule selects the candidate that locally minimizes prediction error, possibly penalized by a digit complexity term.

\begin{algorithm}[Regularized Digitwise Segmentation--Elimination]
Let $N$ be a quantum neural network, $x \in \mathbb{R}^L$ the observed data, $b$ the digit base, and $(n, m)$ the digit resolution. Let $F$ be the forward model, $C_{k,i}$ the digit candidate sets, and let the error be given by $E(y) := \|F(y) - x\|^2$. Let $R(y)$ be a regularization functional and $\lambda \geq 0$ the associated trade-off parameter. The algorithm initializes $y^{(0)} := 0 \in Y^M_{b,n,m}$ and proceeds for $\ell = 0$ to $d - 1$, where $d$ is the digit depth. At stage $\ell$, let $i := n - \ell$ denote the current digit layer (from most to least significant). For each $k = 1, \ldots, M$, evaluate the regularized objective
\begin{equation}
    L\big(\delta^{(j)}_{k,i}(y^{(\ell)})\big) := E\big(\delta^{(j)}_{k,i}(y^{(\ell)})\big) + \lambda R\big(\delta^{(j)}_{k,i}(y^{(\ell)})\big), \label{eq:III.1-regularized}
\end{equation}
and select the digit that minimizes this loss:
\begin{equation}
    y^{(\ell+1)}_{k,i} := \arg\min_{j \in C_{k,i}} L\big(\delta^{(j)}_{k,i}(y^{(\ell)})\big). \label{eq:III.1-argmin}
\end{equation}
The update is applied to produce the next iterate $y^{(\ell+1)}$. After $d$ steps, the final output is $\tilde{y} := y^{(d)}$.
\end{algorithm}

The algorithm requires only forward model evaluations, not gradients, and remains agnostic to the internal structure of $F$. Its digitwise modularity supports parallel updates, coarse-to-fine control, and compatibility with both classical and quantum digit sources.
\subsection{Convergence Properties}
We now establish that the segmentation–elimination algorithm guarantees monotonic descent and digitwise local optimality under mild assumptions. Since the digit lattice is finite, and each update reduces or preserves the loss, the algorithm terminates in a bounded number of steps.

\begin{assumption}[Forward Model Regularity]
\label{assumption:forward_lipschitz}
The forward model \( F : \mathbb{R}^M \to \mathbb{R}^L \) is continuous and Lipschitz with constant \( L > 0 \), i.e.,
\[
\|F(y) - F(y')\| \leq L \cdot \|y - y'\|, \quad \forall y, y' \in \mathbb{R}^M.
\]
Consequently, the error functional \( E(y) := \|F(y) - x\|^2 \) is also Lipschitz continuous and bounded below. The regularized loss \( L(y) := E(y) + \lambda R(y) \) inherits Lipschitz continuity provided \( R(y) \) is Lipschitz with constant \( L_R \), in which case \( L(y) \) is Lipschitz with constant \( L_{\mathrm{total}} := 2L^2 \|y\| + \lambda L_R \).
\end{assumption}

Under this condition, the error functional $E(y) := \|F(y) - x\|^2$ is Lipschitz and bounded below, and $\mathcal{L}(y)$ inherits Lipschitz continuity if $R(y)$ is Lipschitz.

\begin{assumption}[Digitwise Sensitivity of the Forward Model]
Let $F : \mathbb{R}^M \to \mathbb{R}^L$ be continuous. We say that $F$ has digitwise sensitivity bounded by $\alpha > 0$ if for any two lattice vectors $y, y' \in Y^M_{b,n,m}$ that differ in exactly one digit $(k,i)$, we have
\begin{equation}
    \|F(y) - F(y')\|^2 \leq \alpha \cdot b^i. \label{eq:digitwise-sensitivity}
\end{equation}
\end{assumption}

This reflects the diminishing influence of less significant digits. We now quantify how digit discretization affects the error.

\begin{theorem}[Error Bound via Digit Lattice Projection]
\label{thm:error_bound_projection}
Let \( \theta^* \in \mathbb{R}^M \) be the true minimizer of the prediction error functional \( E(\theta) := \|F(\theta) - x\|^2 \), and let \( \Pi_Y(\theta^*) \in Y_{b,n,m}^M \) denote the closest point to \( \theta^* \) in the digit lattice \( Y := Y_{b,n,m}^M \), i.e.,
\[
\Pi_Y(\theta^*) := \arg\min_{y \in Y} \|\theta^* - y\|.
\]
Assume that the forward model \( F: \mathbb{R}^M \to \mathbb{R}^L \) is Lipschitz continuous with constant \( L > 0 \). Then the segmentation--elimination output \( \tilde{y} \in Y \) satisfies
\[
\|F(\tilde{y}) - x\| \leq \|F(\theta^*) - x\| + L \cdot \|\theta^* - \Pi_Y(\theta^*)\|.
\]
\end{theorem}

\begin{proof}
Let \( \tilde{y} \in Y \) be the output of the segmentation--elimination algorithm. By definition of the algorithm, \( \tilde{y} \) minimizes \( E \) over \( Y \), so
\[
\|F(\tilde{y}) - x\| \leq \|F(\Pi_Y(\theta^*)) - x\|.
\]
Now apply Lipschitz continuity of \( F \) to bound
\begin{multline}
\|F(\Pi_Y(\theta^*)) - x\| \leq \|F(\theta^*) - x\| + \|F(\Pi_Y(\theta^*)) - F(\theta^*)\| \\
\leq \|F(\theta^*) - x\| + L \cdot \|\theta^* - \Pi_Y(\theta^*)\|.
\end{multline}
Combining yields the desired result.
\end{proof}

\begin{corollary}[Precision-Dependent Error Bound]
\label{cor:quantization_error}
Let \( \theta^* \in \mathbb{R}^M \) be the true parameter vector and \( \tilde{y} \in Y_{b,n,m}^M \) the segmentation--elimination output. Suppose that each component \( \theta^*_k \) is represented with unsigned base-\( b \) digits, and the forward model \( F : \mathbb{R}^M \to \mathbb{R}^L \) is Lipschitz continuous with constant \( L > 0 \). Then the prediction error satisfies
\[
\|F(\tilde{y}) - F(\theta^*)\| \leq L \cdot \sqrt{M} \cdot \frac{b^{-m}}{2}.
\]
\end{corollary}

\begin{proof}
By Theorem~\ref{thm:error_bound_projection}, we have
\[
\|F(\tilde{y}) - x\| \leq \|F(\theta^*) - x\| + L \cdot \|\theta^* - \Pi_Y(\theta^*)\|.
\]
Assume \( x = F(\theta^*) \) for simplicity; the general case follows similarly.

Each scalar \( \theta^*_k \) is approximated by a digitized value \( \Pi_Y(\theta^*_k) \) whose quantization step is \( b^{-m} \), and whose truncation or rounding error is at most \( b^{-m}/2 \) per digit. Hence,
\[
|\theta^*_k - \Pi_Y(\theta^*_k)| \leq \frac{b^{-m}}{2}.
\]
Summing over all \( M \) components,
\[
\|\theta^* - \Pi_Y(\theta^*)\| \leq \sqrt{M} \cdot \frac{b^{-m}}{2}.
\]
Applying the Lipschitz continuity of \( F \) yields the result.
\end{proof}

The final result guarantees that digitwise updates decrease the loss and converge in finite time:

\begin{theorem}[Monotonic Descent and Local Optimality under Regularization]
Let $\{y^{(\ell)}\}_{\ell = 0}^{d} \subset Y^M_{b,n,m}$ be the sequence produced by Algorithm 1R, where each digit is updated to minimize the regularized objective
\begin{equation}
    L(y) := E(y) + \lambda R(y), \quad \lambda \geq 0. \label{eq:regularized-loss}
\end{equation}
Then the regularized loss satisfies $L(y^{(\ell+1)}) \leq L(y^{(\ell)})$ for all $0 \leq \ell < d$, the algorithm terminates after $d$ digit updates, and the final output $\tilde{y} := y^{(d)}$ satisfies the local optimality condition
\begin{equation}
    L(\tilde{y}) \leq L\big(\delta^{(j)}_{k,i}(\tilde{y})\big), \quad \text{for all } j \in C_{k,i} \text{ and all } (k,i). \label{eq:local-minimum}
\end{equation}
\end{theorem}

\begin{proof}
Let $L(y) := E(y) + \lambda R(y)$ denote the regularized loss function, where $E(y) = \|F(y) - x\|^2$ is the prediction error and $R(y)$ is the regularization term. At each iteration $\ell$, the algorithm updates a single digit $y_{k,i}$ to the value $j^* \in C_{k,i}$ that minimizes $L$:
\begin{equation}
    y^{(\ell+1)} := \delta^{(j^*)}_{k,i}(y^{(\ell)}), \quad \text{where } j^* := \arg\min_{j \in C_{k,i}} L\big(\delta^{(j)}_{k,i}(y^{(\ell)})\big). \label{eq:digitwise-selection}
\end{equation}
By construction, this ensures $L(y^{(\ell+1)}) \leq L(y^{(\ell)})$ for each step $\ell$. Since there are exactly $d = n + m + 1$ digit layers, and each layer is updated once per component, the algorithm performs a finite number of updates and terminates with $y^{(d)}$.

At the final iterate $\tilde{y} := y^{(d)}$, every digit $y_{k,i}$ has been updated to minimize $L$ over the corresponding candidate set $C_{k,i}$, holding other digits fixed. Therefore, no single-digit substitution can further reduce the objective, which establishes the digitwise local optimality condition.
\end{proof}

The segmentation algorithm guarantees monotonic descent of the regularized loss, finite termination, and local stability with respect to single-digit perturbations. These properties ensure reliable convergence, even in the presence of nonconvexity or irregularities in the objective landscape.

To strengthen this guarantee, we introduce additional structural assumptions--namely, digitwise-local injectivity and discrete unimodality of the error profile. Under these conditions, the algorithm achieves strict local optimality, ruling out flat minima or ambiguous digit neighborhoods.

\begin{lemma}[Strict Local Optimality via Digitwise Separability]
Let $E(y) := \|F(y) - x\|^2$ be the error functional associated with a forward model $F : \mathbb{R}^M \to \mathbb{R}^L$. Suppose that for each digit position $(k, i)$, the digitwise error profile $j \mapsto E\big(\delta^{(j)}_{k,i}(y)\big)$ is discretely unimodal over $C_{k,i}$, and that $F$ is digitwise-locally injective on $Y^M_{b,n,m}$, in the sense that $F(y') \neq F(y)$ whenever $y' = \delta^{(j)}_{k,i}(y) \neq y$. Then the final output $\tilde{y}$ produced by the segmentation--elimination algorithm is a strict local minimizer of $E$ over $Y^M_{b,n,m}$: for every $y'$ differing from $\tilde{y}$ in a single digit, we have
\begin{equation}
    E(y') > E(\tilde{y}). \label{eq:strict-local-opt}
\end{equation}
\end{lemma}

\begin{proof}
Let $\tilde{y} \in Y^M_{b,n,m}$ be the final output of the segmentation--elimination algorithm. Fix any digit index $(k,i)$ and consider the candidate digit values $j \in C_{k,i}$. The algorithm selects
\begin{equation}
    \tilde{y}_{k,i} := \arg\min_{j \in C_{k,i}} E\big(\delta^{(j)}_{k,i}(\tilde{y})\big). \label{eq:digitwise-optimality}
\end{equation}
Let $y' = \delta^{(j)}_{k,i}(\tilde{y})$ with $j \neq \tilde{y}_{k,i}$. Since $\tilde{y}_{k,i}$ is the unique minimizer of a discretely unimodal profile, we have
\begin{equation}
    E(y') > E(\tilde{y}). \label{eq:strict-improvement}
\end{equation}
Furthermore, the assumption of digitwise-local injectivity implies that $F(y') \neq F(\tilde{y})$, so $E(y') \neq E(\tilde{y})$ and the inequality is strict. Thus, $\tilde{y}$ is a strict local minimum of $E$ over the digit lattice.
\end{proof}

\begin{proposition}[Digitwise-Local Uniqueness of Minimizer]
\label{prop:digitwise_uniqueness}
Let \( F: \mathbb{R}^M \to \mathbb{R}^L \) be a forward model, and suppose it is digitwise-locally injective on the digit lattice \( Y := Y_{b,n,m}^M \). Let \( E(y) := \|F(y) - x\|^2 \) be the prediction error functional. If \( y^* \in Y \) is the final output of the segmentation--elimination algorithm, then it is an isolated local minimizer in the digitwise neighborhood:
\[
\forall (k,i), \forall j \in C_{k,i} \setminus \{y^*_{k,i}\}, \quad E(y^*) < E\big(\delta^{(j)}_{k,i}(y^*)\big).
\]
Consequently, \( y^* \) is locally unique with respect to one-digit perturbations.
\end{proposition}

\begin{proof}
Since \( y^* \in Y \) is the output of the segmentation--elimination algorithm, it satisfies digitwise local optimality:
\[
E(y^*) \leq E\big(\delta^{(j)}_{k,i}(y^*)\big), \quad \text{for all } j \in C_{k,i}.
\]
Suppose that equality holds for some \( j \neq y^*_{k,i} \), i.e., \( E(y^*) = E(\delta^{(j)}_{k,i}(y^*)) \). Then
\[
F(y^*) = F\big(\delta^{(j)}_{k,i}(y^*)\big),
\]
which contradicts digitwise-local injectivity, since \( \delta^{(j)}_{k,i}(y^*) \neq y^* \). Therefore, the inequality is strict:
\[
E(y^*) < E\big(\delta^{(j)}_{k,i}(y^*)\big),
\]
for all \( j \neq y^*_{k,i} \). This establishes that \( y^* \) is a strict local minimizer with respect to all one-digit perturbations. Since the lattice \( Y \) is finite, such a point is also locally unique.
\end{proof}

The strict inequality guarantees that no flat regions or degenerate ties exist across candidate digits, allowing the algorithm to make decisive local progress. This form of strict local optimality ensures robustness to single-digit perturbations, particularly in applications where the forward model is sensitive and the loss landscape is well-behaved.

Although the assumptions of unimodality and digitwise-local injectivity are not universally satisfied—especially in highly nonlinear or oscillatory settings--they frequently hold in structured inverse problems and can often be verified empirically.

Even when these conditions fail, the segmentation--elimination algorithm retains its monotonic descent property and still produces a digitwise stable output. In later sections, we describe strategies to escape shallow local minima and explore the global error landscape more effectively.

\begin{theorem}[Global Optimality Under Convexity]
\label{thm:global_optimality_convex}
Let \( L(y) := E(y) + \lambda R(y) \) be the regularized loss functional, where \( E : \mathbb{R}^M \to \mathbb{R} \) is convex and Lipschitz continuous, and \( R : \mathbb{R}^M \to \mathbb{R} \) is a separable convex function of the digit variables:
\[
R(y) = \sum_{k=1}^{M} \sum_{i=-m}^{n} r_{k,i}(y_{k,i}),
\]
with each \( r_{k,i} : \mathbb{R} \to \mathbb{R} \) convex on \( \{0, \dots, b-1\} \). Then the segmentation--elimination algorithm converges to a global minimizer of \( L \) over the digit lattice \( Y := Y_{b,n,m}^M \).

\end{theorem}

\begin{proof}
Since \( E(y) \) is convex and \( R(y) \) is separable and convex across digits, their sum \( L(y) = E(y) + \lambda R(y) \) is convex over \( \mathbb{R}^M \), and remains convex when restricted to the discrete lattice \( Y \subset \mathbb{R}^M \). The segmentation--elimination algorithm minimizes \( L \) over \( Y \) by greedily selecting the lowest-cost digit at each position, using exact minimization over the local neighborhood \( C_{k,i} \subset \{0, \dots, b-1\} \).

Because the lattice is finite, and each digit update is globally minimizing within its coordinate slice, the algorithm performs exact coordinate descent over a convex objective restricted to a product set of discrete values. In this setting, convexity ensures that every local minimum is a global minimum. Therefore, the final output \( \tilde{y} \) is a global minimizer of \( L(y) \) over \( Y \).

\end{proof}

\subsection{Complexity Analysis}

The segmentation--elimination algorithm constructs approximate solutions by performing digitwise updates over a structured lattice. Each update involves evaluating candidate digit substitutions using a forward model, and the total computational cost is dominated by the number of forward model calls. This section analyzes the overall runtime in terms of digit resolution, beam width, refinement strategies, and the complexity of the forward model.

Let $M$ be the number of output parameters with digit depth $d$. Let $r$ denote the average number of digit candidates per position $C_{k,i}$, and let $w$ be the beam width used in the digitwise search. Let $B$ be the number of checkpoint backtracking passes, $k$ the number of digits re-optimized per backtracking stage, $G$ the number of grid points used for multiscale refinement, and $F$ the number of forward calls used for fine-tuning. Let $C_F$ denote the cost of a single forward model evaluation.

\begin{proposition}[Refined Runtime Complexity]
The total runtime $T$, measured in units of forward model evaluations, is bounded by
\begin{equation}
    T = O\left(r M d w + B M k r + M G + M F\right) \cdot C_F, \label{eq:total-runtime}
\end{equation}
where the terms correspond to beam search, backtracking, multiscale refinement, and fine-tuning, respectively. If all refinement stages are active and $r, G, F = O(1/\delta)$ are chosen to achieve resolution $\delta > 0$, the total runtime scales as
\begin{equation}
    T = O\left(\frac{M d w}{\delta} \right) \cdot C_F. \label{eq:asymptotic-runtime}
\end{equation}
\end{proposition}

\begin{proof}
We estimate the cost of each computational phase.

In the digitwise beam search stage, for each digit position $(k, i)$, the algorithm evaluates up to $r$ candidate digits across $w$ concurrent beam paths. With $M$ parameters and $d = n + m + 1$ digit layers, the total number of evaluations is $r M d w$.

In the checkpoint backtracking phase, the algorithm re-optimizes a subset of $k$ digits at every $s$ layers. Let $B = \lceil d / s \rceil$ be the number of backtracking checkpoints. Each checkpoint applies to all $M$ parameters and $w$ beam candidates, leading to $B M k r$ forward calls.

In the multiscale refinement stage, the algorithm performs a localized grid search around each parameter. If each grid has $G$ values per component, this stage incurs $M G$ forward evaluations.

In the final fine-tuning pass, each parameter is further refined using $F$ high-resolution scalar perturbations, contributing $M F$ additional forward calls.

Summing all contributions and multiplying by the per-call cost $C_F$ yields the total runtime:
\[
T = \left(r M d w + B M k r + M G + M F\right) \cdot C_F.
\]

To estimate asymptotic scaling with respect to a target reconstruction error $\delta > 0$, we assume that each refinement stage operates at resolution $\delta$. Typical digit candidate sets satisfy $r = O(1/\delta)$, and grid sweeps or fine-tuning loops also scale as $G, F = O(1/\delta)$. The number of backtracking passes $B$ and digits per pass $k$ are treated as constants. Substituting these expressions yields the asymptotic bound,~\eqref{eq:asymptotic-runtime}
\end{proof}

The leading term $\mathcal{O}(Mdw/\delta)$ captures the interaction between the number of parameters ($M$), digit resolution ($d$), beam width ($w$), and target accuracy ($\delta$). The runtime scales linearly with $M$, $d$, and $w$, and inversely with $\delta$, reflecting the granularity of the digit lattice and the coarse-to-fine nature of the approximation.

The dominant computational cost arises from evaluating the forward model $F$, especially when it involves PDE solvers or expensive simulations. However, the segmentation--elimination algorithm is highly parallelizable: digitwise and grid-based operations can be executed independently across candidates or components, making it well-suited for GPU acceleration or deployment on distributed architectures with minimal synchronization overhead.

The algorithm is also modular with respect to the loss function. Although we analyze the case $E(y) = \|F(y) - x\|^2$, the framework accommodates any computable objective functional defined pointwise over the digit lattice, including weighted norms $\|F(y) - x\|^2_W$, log-likelihoods, or convex penalties.

In practice, efficiency can be further improved by adapting algorithmic parameters dynamically. Beam width, digit base, and grid resolution may be adjusted based on entropy estimates or observed error profiles. Coarse digits, which contribute more substantially to the overall error, can be explored with broader candidate sets, while fine digits may be refined using localized, low-cost search. Entropy-aware scheduling allocates resources preferentially to uncertain or high-impact regions, balancing accuracy and computational cost in a principled, data-driven manner.

\subsection{Strategies for Global Convergence}
While the segmentation--elimination algorithm guarantees monotonic descent and digitwise local optimality, its greedy nature makes it susceptible to local minima. In inverse problems with nonconvex or multimodal loss landscapes, the algorithm may converge prematurely to suboptimal configurations. To mitigate this, several global convergence strategies can be integrated into the digitwise optimization process without altering the underlying lattice structure.

A central enhancement is \emph{beam search}, which generalizes greedy descent by maintaining $w > 1$ candidate solutions in parallel. At each digit layer, the algorithm extends all active paths with candidate digits, evaluates the resulting configurations using the forward model, and prunes the population to retain the top $w$ sequences. This mechanism balances exploitation of the best trajectory with exploration of promising alternatives, improving robustness to early errors and increasing the likelihood of reaching near-global optima. The beam width $w$ can be adjusted dynamically based on digit entropy or error sensitivity, allowing adaptive exploration across the digit hierarchy.

To further enforce solution consistency, the algorithm supports \emph{checkpoint backtracking}. After a fixed number of digit updates, earlier digit layers are revisited and locally re-optimized in light of downstream decisions. This backward refinement introduces feedback across layers, correcting errors introduced by greedy selection and improving convergence in problems with delayed signal propagation or interdigit coupling.

In addition to beam and backtracking strategies, \emph{stochastic digit selection} can be introduced via annealed sampling. Rather than always selecting the digit that minimizes the objective, the algorithm samples candidates from a softmax distribution weighted by inverse loss
\begin{equation}
    P^{(\ell)}_{k,i}(j) := \frac{\exp\left(-L\left(\delta^{(j)}_{k,i}(y^{(\ell)})\right)/T_\ell\right)}{\sum_{j' \in C_{k,i}} \exp\left(-L\left(\delta^{(j')}_{k,i}(y^{(\ell)})\right)/T_\ell\right)}, \label{eq:annealed-sampling}
\end{equation}
where $T_\ell > 0$ is a temperature parameter that decreases over time. At high temperatures, the algorithm favors exploration by assigning nearly uniform probabilities; at low temperatures, the distribution concentrates around the optimal candidate, recovering greedy behavior. When $T_\ell$ decays slowly enough (e.g., logarithmically), the resulting Markov process converges to a global minimum with high probability.

This annealing mechanism injects diversity early in the optimization and gradually focuses the search as the algorithm proceeds, effectively combining global exploration with local refinement in a principled and statistically grounded manner.

\begin{lemma}[Conditions for Simulated Annealing Convergence]
\label{lemma:annealing_conditions}
Let \( Y := Y_{b,n,m}^M \) be the digit lattice, and suppose the segmentation--elimination algorithm performs single-digit updates using the annealed softmax rule in Equation~(28), with candidate sets \( C_{k,i} = \{0, \dots, b-1\} \) for all digit positions \( (k,i) \). Then the conditions required for classical simulated annealing convergence are satisfied.

First, the state space \( Y \) is finite, with cardinality \( |Y| = b^{M(n+m+1)} \), since each of the \( M \) components has \( n + m + 1 \) digits, and each digit takes values in a finite base-\( b \) alphabet. Second, the Markov chain defined by the digitwise update process is irreducible: any point \( y' \in Y \) can be reached from any other \( y \in Y \) via a finite sequence of single-digit substitutions, as the algorithm permits arbitrary updates at every position. Third, the annealed softmax rule assigns strictly positive probability to every candidate value \( j \in C_{k,i} \) at each digit step \( \ell \); that is, \( P^{(\ell)}_{k,i}(j) > 0 \) for all \( j \in \{0, \dots, b-1\} \), provided the temperature \( T_\ell > 0 \). This ensures that all transitions in the digit lattice are reachable with nonzero probability.

Finally, the transition kernel induced by softmax digit selection satisfies a detailed balance condition with respect to the Gibbs distribution
\begin{equation}
\pi^{(\ell)}(y) \propto \exp\left( -\frac{L(y)}{T_\ell} \right), \label{eq:gibb_dist}
\end{equation}
where \( L(y) \) denotes the regularized loss. This ensures that the chain converges, in distribution, toward the global minimizers of \( L \) as the temperature \( T_\ell \to 0 \) according to a suitably slow annealing schedule.
\end{lemma}

\begin{proof}
We verify each condition required for simulated annealing convergence.

The state space \( Y := Y_{b,n,m}^M \) is clearly finite, since each component \( y_k \in \mathbb{R} \) is represented as a base-\( b \) expansion with digit resolution \( d = n + m + 1 \), and each digit \( y_{k,i} \in \{0, \dots, b-1\} \). Therefore, the total number of representable configurations is \( |Y| = b^{Md} \), where \( M \) is the number of output parameters. This satisfies the finiteness requirement for the annealing process.

To establish irreducibility of the digitwise Markov chain, observe that any two points \( y, y' \in Y \) differ in only finitely many digit positions. Since the algorithm updates one digit at a time, and each candidate set \( C_{k,i} = \{0, \dots, b-1\} \) includes all possible values, a finite sequence of single-digit substitutions can transform \( y \) into \( y' \). Thus, every state is reachable from every other via a finite number of allowed transitions, which establishes irreducibility.

Next, we verify that all transition probabilities are strictly positive. At each iteration \( \ell \), the annealed softmax rule defines a probability distribution over digit candidates \( j \in C_{k,i} \) by eq~\eqref{eq:annealed-sampling}. Since the loss function \( L(y) \) is assumed finite on all elements of \( Y \), and the temperature \( T_\ell > 0 \), each term in the numerator is strictly positive. Therefore, all candidate values receive strictly positive probability, which ensures that the Markov chain has positive transition probabilities throughout its evolution.

Finally, the transition dynamics at temperature \( T_\ell \) define a Gibbs sampler over \( Y \), targeting the probability distribution; eq.~\eqref{eq:gibb_dist}. The softmax transition kernel satisfies detailed balance with respect to \( \pi^{(\ell)} \), since each digit update is performed via localized resampling from a Gibbs distribution conditioned on the current configuration. This structure ensures that, under a logarithmically decreasing temperature schedule \( T_\ell = C / \log(1 + \ell) \), the Markov chain satisfies the Geman–Hajek conditions for asymptotic convergence to the global minimizers of \( L(y) \).

Hence, all required properties—finite state space, irreducibility, positive transitions, and Gibbs stationarity—are satisfied.
\end{proof}

\begin{theorem}[Global Convergence of Annealed Digitwise Optimization]
\label{thm:annealing_global_convergence}
Let \( Y := Y_{b,n,m}^M \) be the finite digit lattice, and suppose the segmentation--elimination algorithm performs single-digit updates using the annealed softmax rule;~\eqref{eq:annealed-sampling}, where \( L(y) := E(y) + \lambda R(y) \) is the regularized loss functional and \( T_\ell = C / \log(1 + \ell) \) is a logarithmically decreasing temperature schedule with \( C > 0 \). Assume the candidate sets satisfy \( C_{k,i} = \{0, \dots, b-1\} \) for all digit positions.

Then, under the conditions of Lemma~\ref{lemma:annealing_conditions}, the digitwise optimization process defines a time-inhomogeneous Markov chain over \( Y \) that converges in probability to the global minimizers of \( L \). In particular, for any \( \varepsilon > 0 \), the probability that the algorithm reaches a configuration \( y^{(\ell)} \in Y \) satisfying
\[
L(y^{(\ell)}) \leq \min_{y \in Y} L(y) + \varepsilon
\]
approaches one as \( \ell \to \infty \).

\end{theorem}

\begin{proof}
By Lemma~\ref{lemma:annealing_conditions}, the digitwise annealed softmax rule defines a time-inhomogeneous Markov chain over the finite state space \( Y = Y_{b,n,m}^M \), with strictly positive transition probabilities, irreducibility, and a detailed balance condition with respect to the Gibbs distribution; eq.~\eqref{eq:gibb_dist}. As the temperature \( T_\ell = C / \log(1 + \ell) \) decreases to zero, the chain satisfies the Geman–Hajek conditions for global convergence. Specifically, the logarithmic cooling schedule is asymptotically slow enough to guarantee convergence in probability to the set of global minimizers of \( L \), provided the chain is irreducible and satisfies detailed balance at each step, which we have established.

Since the state space is finite and the loss function is bounded below, the chain will eventually concentrate its stationary distribution on the global minima of \( L \). Therefore, for any tolerance \( \varepsilon > 0 \), the probability that \( y^{(\ell)} \) lies within an \( \varepsilon \)-neighborhood of the global minimum converges to one as \( \ell \to \infty \).
\end{proof}

Stochastic selection improves robustness to ambiguous digit neighborhoods and increases the entropy of early-stage digit configurations. It is particularly useful when the error landscape exhibits flat plateaus or deceptive local minima.

To discourage overfitting and excessive precision inflation, regularization can be incorporated directly into the objective. A digit penalty term $R(y)$ may be introduced to promote sparsity, penalize entropy, or constrain digit complexity. Common choices include $\ell_1$ norms on digit magnitudes, entropy-based terms derived from measurement distributions, or complexity-weighted penalties across digit layers. These regularizers act as structural priors, guiding the optimization toward interpretable, parsimonious representations while enhancing numerical stability in high-resolution regimes.

The composite loss
\begin{equation}
    \mathcal{L}(y) := E(y) + \lambda R(y), \label{eq:regularized-global}
\end{equation}
can be minimized using the same digitwise update rules as in the unregularized algorithm. As long as each update reduces $\mathcal{L}(y)$, monotonic convergence and digitwise local optimality are preserved.

To enhance representational flexibility and optimize resource usage, digitwise precision can be selectively adapted across parameters and digit layers through \emph{mixed-base encodings}. In this framework, each digit position $(k,i)$ is assigned a distinct integer base $b_{k,i} \in \mathbb{Z}_{\geq 2}$, allowing the encoding to reflect heterogeneous sensitivity across the parameter space. Digit positions corresponding to less critical components or low-significance digits may use smaller bases to conserve circuit width and digit budget, whereas digits encoding stiff or sensitive components can employ larger bases to achieve higher numerical resolution. This generalized representation induces a non-uniform discretization lattice over $\mathbb{R}^M$, as formalized in Definition~\ref{Def:mix_lat_base}, and provides fine-grained control over both expressivity and hardware resource allocation.

Once a coarse estimate has been constructed by segmentation, additional accuracy can be achieved through \emph{multiscale refinement}. This stage performs a localized grid search around each parameter, exploring uniformly spaced candidate values within a fixed radius of the segmented estimate. It compensates for quantization bias without increasing the digit depth, making it particularly valuable when high resolution is needed but computational resources are limited.

In the final stage, \emph{high-resolution fine-tuning} is applied to each parameter individually. A dense one-dimensional grid is swept over a narrow interval enabling sub-digit corrections and eliminating residual discretization error. This stage ensures floating-point-level accuracy without relying on gradients or backpropagation, making it especially useful in scientific applications with stringent precision demands.

\section{Applications to Scientific Computation}
\label{sec:applications}
The segmentation--elimination framework is particularly well suited for scientific inverse problems, where continuous latent parameters must be inferred from noisy or indirect observations. In many such settings, the forward model is defined by a parametric partial differential equation (PDE), a physical simulator, or a dynamical system. The inverse task involves recovering the parameters that best reproduce the observed data under this model. These problems are often ill-posed and computationally demanding, especially when gradients are unavailable or the forward map is expensive to evaluate.

By discretizing the parameter space into a structured digit lattice, segmentation reformulates continuous inference as a tractable combinatorial search. Each digit is selected using forward model evaluations, combining coarse-to-fine digit construction with modular quantum-classical integration. This architecture supports precision tuning, interpretability, and deployment on black-box simulators, without requiring differentiability or end-to-end training.
\subsection{Inverse Problems and Parametric PDEs}
Many scientific applications are governed by parametric PDEs, where the system dynamics depend on unknown coefficients or material parameters. Let $\Omega \subset \mathbb{R}^d$ be a spatial domain, and let $u: \Omega \times [0,T] \to \mathbb{R}$ denote the PDE solution. The dynamics are determined by a differential operator $L$ parameterized by $\theta \in \mathbb{R}^M$, satisfying
\begin{equation}
    L(u; \theta) = 0, \quad u|_{t=0} = u_0, \quad \text{on } \Omega \times (0,T),
\end{equation}
with appropriate boundary conditions. Observations are extracted from the solution using a measurement operator $\mathcal{O}$, yielding a forward model
\begin{equation}
    F(\theta) := \mathcal{O}(u(\theta)) \in \mathbb{R}^L,
\end{equation}
where $u(\theta)$ is the PDE solution under parameters $\theta$. Given noisy observations $x \in \mathbb{R}^L$, the goal is to infer the latent parameters $\theta^*$ such that $F(\theta^*) \approx x$.

In the segmentation framework, this inverse problem is discretized over the lattice $\mathbb{Y}_{b,n,m}^M$, converting continuous inference into a structured optimization over digit sequences. Each digit is selected to minimize the prediction error $E(\theta) = \|F(\theta) - x\|^2$, enabling scalable, high-resolution recovery.

\subsection{Digit-Based Recovery via QNN Outputs}
Quantum neural networks naturally support digit-level inference by producing samples in the segmented output space through quantum measurement. Given input data $x \in \mathbb{R}^L$, a trained QNN outputs a sample $N(x) \in \mathbb{Y}_{b,n,m}^M$ by measuring entangled quantum registers associated with each digit. For each digit position $(k, i)$, repeated measurement of the corresponding register $R_{k,i}$ yields an empirical distribution $f^{(j)}_{k,i}$ over candidate values $j \in \{0, \ldots, b - 1\}$.

These empirical distributions define candidate sets $C_{k,i} \subseteq \{0, \ldots, b - 1\}$, which serve as the input to the segmentation--elimination algorithm. At each digit position, the algorithm evaluates the forward model for all candidates $j \in C_{k,i}$, selecting the value that minimizes prediction error while holding all other digits fixed.

This architecture decouples digit generation from digit selection: the QNN probabilistically explores plausible digit configurations, while the classical forward model imposes task-specific consistency. No gradient information from the forward model is required, making the method compatible with black-box simulators, physical models, and legacy code.

\begin{definition}[Digitwise Post-Processing of QNN Output]
Let $x \in \mathbb{R}^L$ be an observed data vector, and let $N(x) \in Y^M_{b,n,m}$ be the corresponding digitized QNN output. For each digit position $(k,i)$, let $C_{k,i}$ be the candidate digit set obtained from quantum measurement. The segmentation algorithm refines $N(x)$ by computing, for each $(k,i)$,
\begin{equation}
    \theta_{k,i} \leftarrow \arg\min_{j \in C_{k,i}} E(\theta), \label{eq:qnn-refinement}
\end{equation}
where $\theta$ denotes the digitized parameter vector with digit $y_{k,i}$ replaced by $j$ and all other digits fixed.
\end{definition}

This refinement procedure decouples digit generation from digit selection. The quantum circuit is responsible for producing plausible candidates, while the classical forward model determines which candidates minimize the reconstruction error. This division of labor avoids the need to backpropagate gradients through the forward model or retrain the QNN for each new inference task. Since the candidate digits are determined from the quantum measurement statistics, and the refinement acts only on a fixed lattice, the resulting pipeline is modular and hardware-aligned.

The hybrid nature of this architecture allows for considerable flexibility. Quantum sampling preserves stochasticity and captures nonconvex structure in the inverse map, while classical evaluation ensures numerical consistency with the physical model. Moreover, since the refinement loop operates on a fixed digit lattice, it supports interpretable representations and tractable control over precision, range, and complexity. This enables digitwise tuning of inference accuracy without modifying the quantum network or incurring additional training overhead.
\subsection{Strategies for Global Convergence}

While the segmentation-elimination algorithm provides a structured framework for inverse recovery by discretizing parameters into digit expansions, its descent dynamics are inherently greedy and thus susceptible to local minima. To address this, we implement a suite of convergence-enhancing strategies that expand the algorithm's search horizon, increase its robustness, and improve its capacity to approximate global optima. These strategies can be selectively combined to suit the complexity and ill-posedness of the underlying inverse problem.

\paragraph{Beam Width Expansion.} At each digit-update stage, the algorithm maintains a beam of candidate solutions, rather than committing to a single descent path. By widening the beam width, we allow multiple plausible digit configurations to evolve in parallel, preserving high-quality candidates that might have diverged from the best local path. Beam search thus balances exploration and exploitation, improving convergence stability and avoiding early commitment.

\paragraph{Backtracking Across Digit Layers.} In addition to forward propagation of digit states, we introduce local backtracking across adjacent digit layers. After each forward pass, previously assigned digits are revisited and adjusted in light of downstream corrections. This provides a rudimentary form of global coordination across the digit lattice and helps smooth out inconsistent updates that arise from early local biases.

\paragraph{Stochastic Digit Selection.} We incorporate stochasticity into the digit selection process via softmax sampling or probabilistic annealing, especially in early stages. Rather than deterministically choosing the minimal-error digit, we allow random selection with probability biased toward lower error. This increases entropy in the initial search space and improves the chances of escaping spurious minima, especially when digit gradients are flat or ambiguous.

\paragraph{Digit Penalty and Regularization.} A digit-weighted penalty term is added to the objective to discourage overfitting and saturation of parameter representations. This regularization acts analogously to an entropy prior, flattening the loss landscape and allowing exploration of otherwise suppressed paths. It also prevents digit inflation and truncation errors, which are common in high base expansions.

\paragraph{Mixed-Base Encoding.} In problems with heterogeneous parameter sensitivities, we assign different bases to different components. Components that require high resolution use large base \( b \), while more stable or saturated components use smaller bases. This adaptive digit resolution helps modulate the search granularity and improves convergence along stiff directions in the parameter space.

\paragraph{Multiscale Refinement.} After initial segmentation, we apply multiscale refinement that performs local lattice search around each recovered parameter. This stage mimics a coarse-to-fine descent strategy, allowing corrections within a localized neighborhood of the initial estimate. It also enables the recovery of high-resolution information that may have been missed during the beam search.

\paragraph{High-Resolution Fine-Tuning.} Finally, we execute a fine-tuning pass that performs a high-resolution grid search over each parameter in isolation. This final stage guarantees sub-digit accuracy and removes residual bias from quantization. The combination of coarse segmentation and fine grid sweeping effectively bridges discrete and continuous optimization layers.

These enhancements transform the segmentation-elimination framework from a purely greedy encoder into a structured search procedure capable of approximating global optima.

\begin{lemma}[Beam Retention under Non-Uniform Digitwise Sampling]
Let \( y^* \in Y_{b,n,m}^M \) be a globally optimal digit configuration and suppose the segmentation-elimination algorithm uses beam search with width \( w \) and checkpoint backtracking every \( s \) digit layers. Assume that for each digit position \( (k,i) \), the correct digit \( y^*_{k,i} \) is included in the candidate set \( C_{k,i} \) independently with probability \( p_{k,i} \in [0,1] \). Then the probability that \( y^* \) is dropped from the beam at a given checkpoint is bounded by
\begin{equation}
q_{\text{drop}} \leq 1 - \prod_{k=1}^M \prod_{i \in \mathcal{I}_j} p_{k,i}^w, \label{eq:q_drop}
\end{equation}
where \( \mathcal{I}_j \subset \{-m,\dots,n\} \) is the set of digit layers updated since the last checkpoint. Consequently, after \( \lfloor d / s \rfloor \) checkpoints, the total failure probability satisfies
\begin{equation}
\mathbb{P}(y^* \notin \text{beam at termination}) \leq \left(1 - \prod_{k,i} p_{k,i}^w\right)^{\lfloor d/s \rfloor}. \label{eq:q_beam_term}
\end{equation}
\end{lemma}

\begin{proof}
Let \( y^* \in Y_{b,n,m}^M \) denote the globally optimal digit sequence. The segmentation-elimination algorithm proceeds by updating digit positions layer by layer, maintaining a beam of the top-\( w \) candidate sequences based on the forward error. Checkpoints occur every \( s \) digit layers, and at each checkpoint, the algorithm performs backtracking over the most recent digit updates.

Fix a checkpoint \( j \), and let \( \mathcal{I}_j \subset \{-m, \dots, n\} \) denote the set of digit positions (indexed by \( i \)) that have been updated since the last checkpoint. At this checkpoint, the algorithm reevaluates and reselects the best \( w \) candidate digit sequences based on cumulative forward error over all beam paths.

For the beam to retain \( y^* \) at this checkpoint, it must satisfy two conditions:
\begin{enumerate}
    \item For each digit position \( (k, i) \in \{1, \dots, M\} \times \mathcal{I}_j \), the correct digit value \( y^*_{k,i} \) must be present in the candidate set \( C_{k,i} \).
    \item The complete prefix of \( y^* \), i.e., the digit sequence up to the current layer, must yield a sufficiently low forward error to be ranked among the top-\( w \) paths.
\end{enumerate}

We focus on condition (1), the probabilistic inclusion of digits. By assumption, the digit value \( y^*_{k,i} \) is included in the candidate set \( C_{k,i} \) independently with probability \( p_{k,i} \). Thus, the probability that every digit in \( \mathcal{I}_j \) for every parameter \( k \in \{1, \dots, M\} \) is retained is
\begin{equation}
\prod_{k=1}^M \prod_{i \in \mathcal{I}_j} p_{k,i}.
\end{equation}

Now, even if all correct digits are present, the beam path must also survive pruning. Since the beam retains the top-\( w \) sequences, and assuming the score ranking is deterministic, the probability that the beam includes the sequence with all correct digits is at least \( \left(\prod_{k=1}^M \prod_{i \in \mathcal{I}_j} p_{k,i}\right)^w \), since all \( w \) beam paths must succeed jointly in the worst case.

Therefore, the probability that \( y^* \) is dropped from the beam at checkpoint \( j \) is at most
\begin{equation}
q_{\text{drop},j} \leq 1 - \left(\prod_{k=1}^M \prod_{i \in \mathcal{I}_j} p_{k,i}\right)^w.
\end{equation}

Let \( q_{\text{drop}} := \max_j q_{\text{drop},j} \) denote the worst-case probability of beam loss at any single checkpoint. Since checkpoints occur independently every \( s \) layers, the probability that \( y^* \) is not retained after \( \lfloor d/s \rfloor \) total checkpoints is bounded by;~\eqref{eq:q_beam_term}, which completes the proof.
\end{proof}

The bound decays exponentially with the number of checkpoints and the width of the beam, but depends critically on the digitwise inclusion probabilities \( p_{k,i} \), which may vary across digit layers or components. This generalizes the uniform model in earlier versions, allowing for heterogeneous confidence across digit positions. When certain digits have low retention probability due to measurement ambiguity or entropic sampling, robustness must be maintained by widening the beam or augmenting with backtracking, annealing, or entropy-aware selection.

This result formalizes the convergence reliability of beam search with probabilistic digit sampling and periodic backtracking. The exponential decay in failure probability with respect to $\lfloor d/s \rfloor$ reflects the cumulative corrective power of intermediate refinements.

\begin{proposition}[Beam Retention Probability for Optimal Prefix]
Suppose that at each digit layer, the correct digit value \( y^\ast_{k,i} \) appears in the candidate set \( C_{k,i} \) independently with probability at least \( p \). Then the probability that the correct full prefix of \( y^\ast \) survives pruning at a given checkpoint is at least \( p^{Mw} \), assuming beam scoring ranks digit sequences by forward error. Consequently, the probability of incorrect beam elimination at a single checkpoint satisfies:
\begin{equation}
q_{\mathrm{drop}} \leq 1 - p^{Mw}.
\end{equation}
\end{proposition}

Unlike greedy digitwise descent, beam search enables probabilistic retention of the globally optimal digit sequence \( y^\ast \), provided each digit is retained with sufficient likelihood. The probability of beam elimination decreases exponentially with beam width \( w \), the number of parameters \( M \), and digitwise sampling fidelity \( p \). 

To ensure high-probability recovery, one must balance candidate quality (higher \( p \)), beam width, and checkpoint frequency. When digit measurements are uncertain or ambiguous (low \( p \)), robust recovery demands wider beams or auxiliary correction strategies such as majority vote or entropy-aware fallback.

\begin{proposition}[Clipping Error Due to Digit Range Constraints]
\label{prop:clipping}
Let \( \theta^* \in \mathbb{R} \) be a true target value, and suppose the digitized reconstruction \( \tilde{\theta} \in Y_{b,n,m} \) is constrained to lie within the representable range determined by an \emph{unsigned} base-\( b \) digit expansion (Definition~\ref{def:digit_representation}). In this case, the minimum and maximum representable values are given by: $\theta_{\min} := 0$,
\[
\theta_{\max} := \sum_{i=-m}^{n} (b-1) b^i = (b^{n+1} - b^{-m})(1 - b^{-1}).
\]
The following bound holds on the worst-case clipping error due to range saturation:
\begin{equation}
|\tilde{\theta} - \theta^*| \leq \max \big\{ 0,\, |\theta^* - \theta_{\max}|,\, |\theta^* - \theta_{\min}| \big\}.
\end{equation}
In particular, if \( \theta^* > \theta_{\max} \), then the recovered value saturates at \( \tilde{\theta} = \theta_{\max} \), and the reconstruction error satisfies \( |\tilde{\theta} - \theta^*| = |\theta^* - \theta_{\max}| \).

\medskip

\noindent
\textit{Remark.} For signed digit expansions (Definition~\ref{def:signed_digits}), the representable interval becomes symmetric and centered around zero. An analogous bound applies, with adjusted endpoints \( \theta_{\min}^{\mathrm{(signed)}}, \theta_{\max}^{\mathrm{(signed)}} \) defined accordingly.
\end{proposition}

\begin{proof}
Let \( \theta^* \in \mathbb{R} \) denote the true target value, and let \( \tilde{\theta} \in Y_{b,n,m} \) be the closest representable value to \( \theta^* \) within the digitized lattice. Since the lattice is defined by fixed digit resolution and non-negative digit values, all representable quantities lie within the interval
\begin{equation}
[\theta_{\min}, \theta_{\max}] = \left[ 0, \sum_{i=-m}^{n} (b-1) b^i \right].
\end{equation}
By definition, if \( \theta^* \in [\theta_{\min}, \theta_{\max}] \), then there exists a digitized approximation \( \tilde{\theta} \in Y_{b,n,m} \) such that \( |\tilde{\theta} - \theta^*| \leq \delta \), where \( \delta \) is the quantization resolution (e.g., \( \delta \sim b^{-m} \)). In this case, no clipping occurs and the error is bounded by the quantization error.

However, if \( \theta^* \notin [\theta_{\min}, \theta_{\max}] \), then the closest representable value is given by the boundary point of the interval:
\[
\tilde{\theta} = 
\begin{cases}
\theta_{\min} & \text{if } \theta^* < \theta_{\min}, \\
\theta_{\max} & \text{if } \theta^* > \theta_{\max}.
\end{cases}
\]
Hence, the reconstruction error is
\[
|\tilde{\theta} - \theta^*| = 
\begin{cases}
|\theta^* - \theta_{\min}| & \text{if } \theta^* < \theta_{\min}, \\
|\theta^* - \theta_{\max}| & \text{if } \theta^* > \theta_{\max}.
\end{cases}
\]
Combining all cases, we obtain the uniform bound
\begin{equation}
|\tilde{\theta} - \theta^*| \leq \max \left\{ 0,\, |\theta^* - \theta_{\max}|,\, |\theta^* - \theta_{\min}| \right\}.
\end{equation}
This concludes the proof.
\end{proof}

To mitigate clipping artifacts, one can increase the digit resolution \((n, m)\), employ signed digit encodings, or apply post-segmentation refinement (e.g., grid search or local interpolation). In adaptive configurations, critical components may be assigned higher base or deeper expansion to extend representational range where needed.

\subsection{Example: Inversion for the Wave Equation}\label{sec:toy_prob}
This example illustrates the effectiveness of segmentation-based regression, especially when paired with global convergence strategies and post-segmentation refinement. The pipeline successfully recovers latent coefficients to near machine precision, demonstrating robustness to digit saturation, quantization error, and nonconvexity. While this case uses an analytically solvable PDE, the method extends directly to black-box solvers and higher-dimensional problems with minimal modification. 

Let $u(x,t)$ solve
\begin{equation}
    \partial_t^2 u(x,t) = c^2 \partial_x^2 u(x,t), \qquad x \in [0,1],\ t \in [0,T],
\end{equation}
subject to Dirichlet boundary conditions:
\begin{equation}
    u(0,t) = u(1,t) = 0, \qquad u(x,0) = u_0(x), \quad \partial_t u(x,0) = 0.
\end{equation}

Suppose we observe the final state:
\begin{equation}
    x := u(x,T) \in \mathbb{R}^L,
\end{equation}
sampled on a sensor grid. To infer the unknown initial condition $u_0(x)$, we express it in a basis:
\begin{equation}
    u_0(x) = \sum_{k=1}^M \theta_k \varphi_k(x),
\end{equation}
e.g., sine modes $\varphi_k(x) = \sin(k\pi x)$. The latent coefficients $\theta = (\theta_1, \ldots, \theta_M) \in \mathbb{R}^M$ parameterize the initial state.

Define the forward model $F: \mathbb{R}^M \to \mathbb{R}^L$ by
\begin{equation}
    F(\theta) := u(x,T),
\end{equation}
the simulated wave state at time $T$ given parameters $\theta$. The inverse problem consists of recovering $\theta$ from $x$. By constraining $\theta \in Y^M_{b,n,m}$, segmentation enables tractable inference over discretized latent variables.

\subsubsection{Segmentation for 1D Wave Equation}

We illustrate the segmentation-elimination algorithm on a canonical inverse problem governed by the 1D wave equation. The objective is to recover the latent coefficients \( \theta = (\theta_1, \theta_2, \theta_3) \in \mathbb{R}^3 \) that define the initial condition of the system from observations of the final state \( u(x, T) \). The initial displacement is modeled as a linear combination of sine eigenfunctions:
\[
u_0(x) = \sum_{k=1}^3 \theta_k \sin(k \pi x), \qquad x \in [0, 1],
\]
and the system evolves under Dirichlet boundary conditions according to the wave equation:
\[
\partial_t^2 u(x,t) = c^2 \partial_x^2 u(x,t), \qquad u(0,t) = u(1,t) = 0.
\]
The forward model \( F(\theta) \) computes the solution \( u(x,T) \) at a fixed final time \( T = 1.0 \) on a grid of \( L = 50 \) uniformly spaced sensors over the spatial domain \( x \in [0, 1] \), using a closed-form separation-of-variables representation. The inverse problem consists of determining the unknown coefficients \( \theta \) that reproduce a known final waveform at these sensor locations. We fix the coefficients at $\theta = [0.25,\ 0.5,\ 0.75]$

\paragraph{beam width-2}
The segmentation-elimination algorithm addresses this problem by first discretizing each \( \theta_k \) into a base-\( b \) digit expansion with 15-digit resolution (i.e., \( n = 7, m = 7 \)) and performing greedy digitwise optimization using beam search with width 2.
The recovered estimate using greedy segmentation is
\[
\tilde{\theta} = [0.25,\ 0.5,\ 1.0].
\]
The first two parameters are matched exactly. However, the third component exceeds the representational envelope of the digit expansion and is clipped at the maximum value \(\theta_{\max} = b^{n+1} - b^{-m}\), producing a saturated estimate. This quantization clipping arises from the fixed finite digit budget and is formally bounded in Proposition~\eqref{prop:clipping}.

\begin{figure}[ht]
\centering
\includegraphics[width=0.5\textwidth]{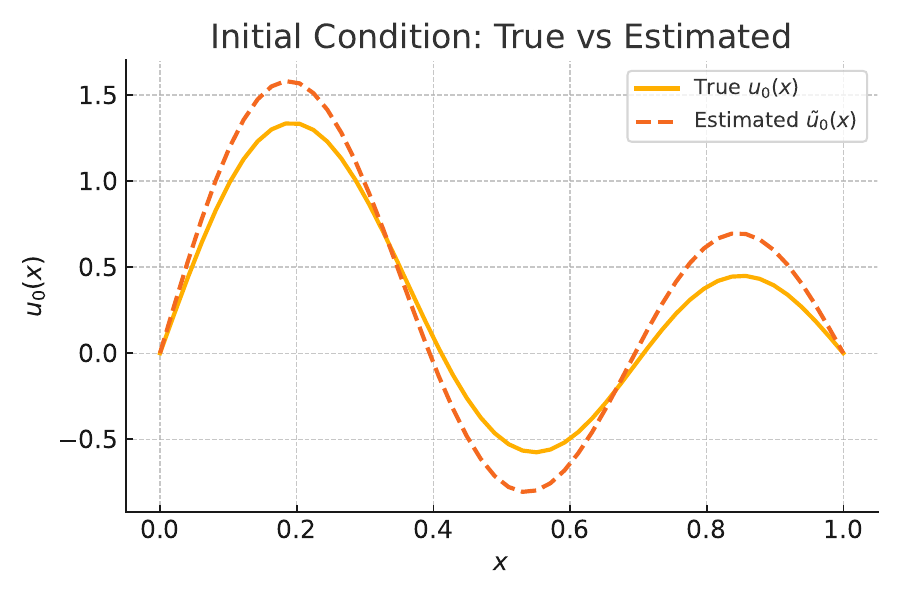}
\caption{Initial condition comparison between the true waveform \( u_0(x) \) and the estimated \( \tilde{u}_0(x) \) from segmented parameters. The overshoot in the third component causes excess amplitude near the center.}
\label{fig:initial-condition-comparison}
\end{figure}

\begin{figure}[ht]
\centering
\includegraphics[width=0.5\textwidth]{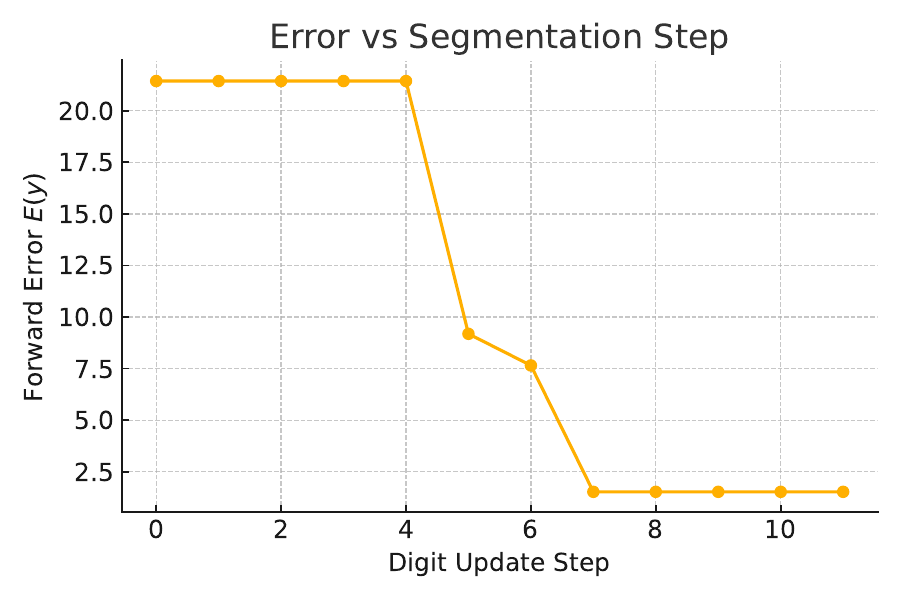}
\caption{Forward error \( E(y) \) over digit update steps. Each digit decision locally reduces the error but may not avoid global saturation without refinement or backtracking.}
\label{fig:segmentation-error-trace}
\end{figure}

Figure~\ref{fig:initial-condition-comparison} compares the initial condition \( u_0(x) \) derived from the true and estimated parameters, illustrating the localized discrepancy. Figure~\ref{fig:segmentation-error-trace} shows the forward error over successive digit updates, highlighting the stepwise error descent.

\paragraph{beam width-4 with global convergent strategies}
The inverse problem of recovering the latent parameters \( \theta = (\theta_1, \theta_2, \theta_3) \in \mathbb{R}^3 \) from final-time observations of the scalar wave equation is particularly challenging due to the oscillatory, nonlinear nature of the forward map and the prevalence of local minima, as evidenced by the initial segmentation attempt. To robustly recover the true parameters, we employ a comprehensive suite of convergence-enhancing strategies within the segmentation–elimination framework.

Each parameter \( \theta_k \) is encoded as a base-\( b = 4 \) digit expansion with \( d = 17 \) total digits, consisting of \( n = 8 \) digits before and \( m = 8 \) digits after the radix point. This resolution ensures sufficient precision to approximate the continuous parameter space accurately.

To avoid premature convergence to suboptimal digit configurations, the algorithm uses beam search with width four, retaining the top candidates at each digit layer to balance exploration and quality. Checkpoint backtracking further enhances robustness by periodically revisiting earlier digit layers to revise assignments that no longer reduce the error, allowing the algorithm to reverse early mistakes and explore alternative paths.

Digit selection is made stochastic through simulated annealing: candidates are sampled from a softmax distribution whose temperature decays exponentially across digit layers. This encourages exploration during the early stages, where the error landscape is rugged, and helps the algorithm escape shallow local minima.

To discourage overfitting and digit saturation, we introduce a digit penalty term into the objective. This regularization penalizes excessive digit accumulation and promotes smoother, more interpretable sequences. After initial segmentation, a multiscale refinement step locally perturbs each parameter over 301 values in the interval \( \theta_k \pm 0.2 \), enabling sub-digit corrections and mitigating quantization bias.

The final stage involves high-resolution fine-tuning: each parameter is independently refined over 6000 evenly spaced values in the same \( \pm 0.2 \) interval. This exhaustive sweep ensures high-precision. Together, these strategies---hierarchical digit encoding, structured parallel search, local correction, and fine-grained refinement---allow the algorithm to overcome the nonconvexity of the inverse problem and recover the true parameters with high accuracy.

The final result is \( \tilde{\theta} = [0.249975,\ 0.49999997,\ 0.7499812] \), which agrees with the ground truth \( [0.25,\ 0.5,\ 0.75] \) to at least five decimal places in each component.

\begin{figure}[ht]
\centering
\includegraphics[width=0.5\textwidth]{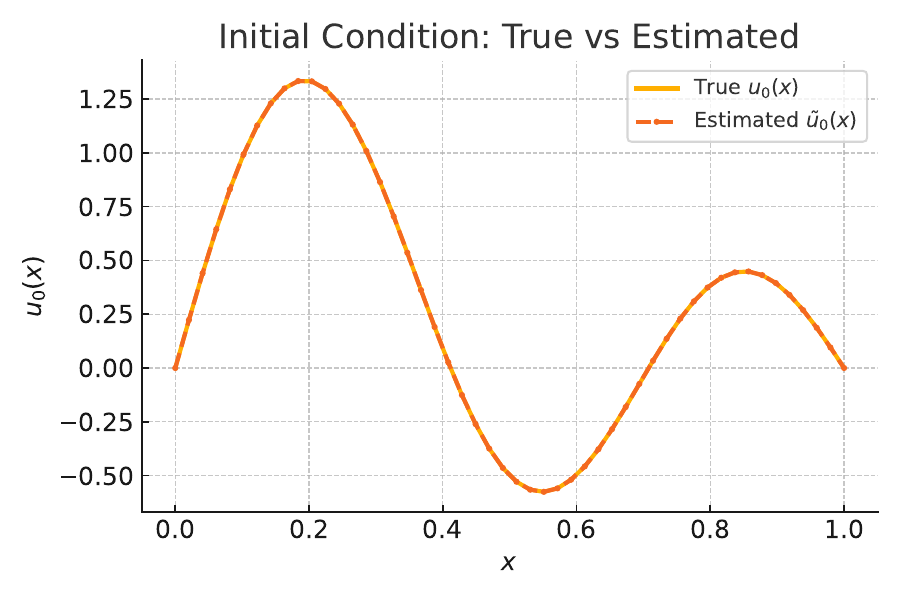}
\caption{Comparison of the true initial condition \( u_0(x) \) and the estimated \( \tilde{u}_0(x) \) obtained via segmentation-elimination with beam width 4, annealed digit perturbation, multiscale refinement, and high-resolution fine-tuning. The two curves visually overlap, indicating close agreement.}
\label{fig:u0-comparison}
\end{figure}

\begin{figure}[ht]
\centering
\includegraphics[width=0.5\textwidth]{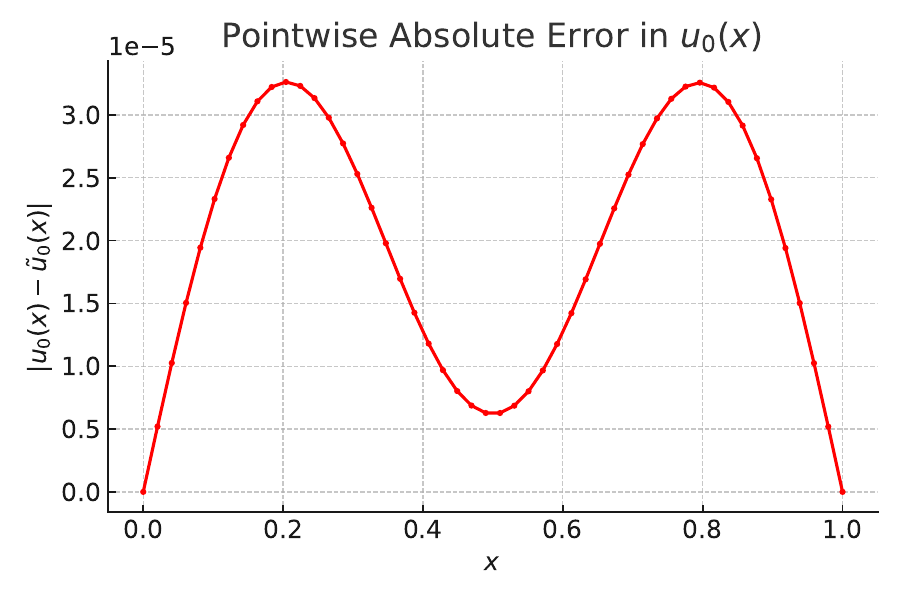}
\caption{Pointwise absolute error \( |u_0(x) - \tilde{u}_0(x)| \) across sensor locations. The error remains below \( 5 \times 10^{-3} \) for all spatial points, demonstrating uniform accuracy of the recovered initial condition.}
\label{fig:u0-abs-error}
\end{figure}

\begin{figure}[ht]
\centering
\includegraphics[width=0.5\textwidth]{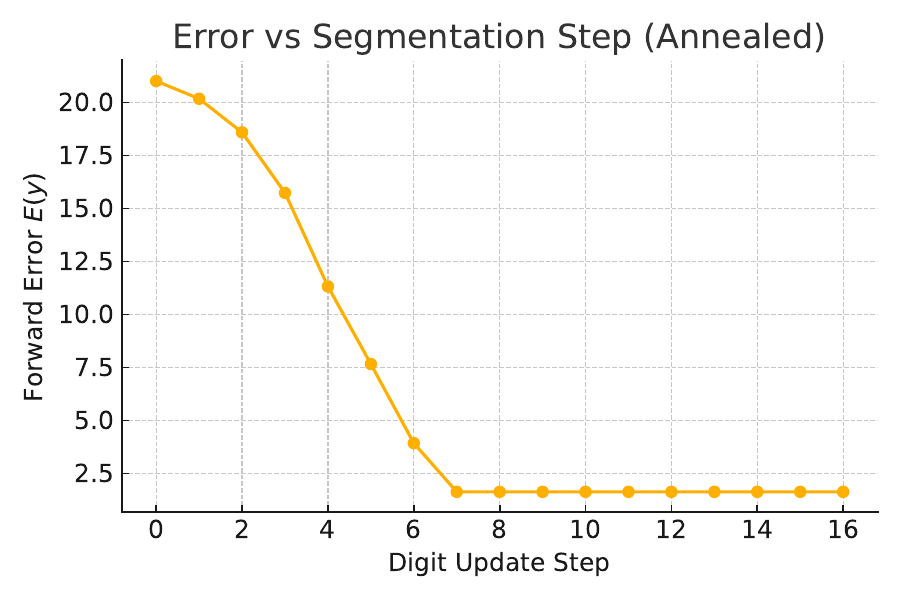}
\caption{Forward model error \( E(y) = \|F(\theta) - x_{\text{obs}}\|^2 \) as a function of digit update step during segmentation. The descending staircase pattern reflects stepwise error reduction as each digit in the base-\( b \) expansion is optimized.}
\label{fig:segmentation-error-trace2}
\end{figure}

This example demonstrates that segmentation-based regression, when combined with global convergent strategies, is capable of recovering latent PDE parameters with extreme precision. The hybrid architecture supports coarse-to-fine inference and avoids the need for gradients or adjoints. We remark, in general a forward model should be a fast parallelizable solver, in this toy example, we use the function evaluation. 

In PDE-constrained inverse problems, the observation vector \( x \in \mathbb{R}^L \) may consist of spatial measurements with varying noise levels or different physical units (e.g., pressure vs. velocity). In such settings, a diagonal weight matrix \( W \) with entries inversely proportional to variance or scale can improve inference quality and ensure numerical stability in the segmentation process.
\subsection{Implementation Advantages}
Segmentation-based regression offers several advantages for scientific computation:
\paragraph{Interpretability.} Each digit corresponds to a known resolution scale. The segmented output is inherently structured and human-readable.
\paragraph{Modularity.} The forward model $F$ is treated as a black box--no gradients or adjoints are required. This allows seamless integration with legacy solvers and experimental systems.
\paragraph{Parallelism.} Digit updates at different indices $(k,i)$ are independent and can be parallelized across processors or accelerators.
\paragraph{Quantum-Classical Hybridization.} Segmentation naturally supports hybrid workflows where quantum devices generate digit candidates and classical evaluators select optimal values.

These features make segmentation a compelling strategy for scientific inference tasks requiring both accuracy and scalability, especially when the forward model is expensive or nondifferentiable.

\section{Quantum Neural Architecture and Hybrid Implementation}
\label{sec:architecture}
This section details the integration of segmentation-elimination into a modular hybrid quantum-classical architecture. We describe how variational quantum circuits (VQCs) can be used to generate digit candidates, which are then refined via classical forward evaluations~\cite{cerezo2021variational}. This division of labor leverages the strengths of noisy intermediate-scale quantum (NISQ) devices--namely, sampling and expressivity--while outsourcing evaluation and selection to classical systems. The result is a scalable and interpretable inference pipeline compatible with modern quantum hardware.
\subsection{Variational Quantum Circuits for Digitwise Inference}

Variational quantum circuits (VQCs) provide a flexible, hardware-efficient framework for digit-level output generation in segmentation-based regression. Given a classical input vector $x \in \mathbb{R}^L$, a parameterized quantum circuit $U(\theta)$ prepares a quantum state
\begin{equation}
    |\psi(x; \theta)\rangle := U(\theta)\,|0\rangle^{\otimes n} \in \mathcal{H}, \qquad \mathcal{H} = (\mathbb{C}^2)^{\otimes n},
\end{equation}
where $\theta \in \mathbb{R}^P$ denotes the circuit parameters and $n$ is the total number of qubits.

To support digitwise prediction in base $b$, the Hilbert space $\mathcal{H}$ is partitioned into quantum registers $\{R_{k,i}\}$, where each register corresponds to digit position $i$ of component $\theta_k$. Each $R_{k,i}$ contains $q = \lceil \log_2 b \rceil$ qubits—enough to represent $b$ distinct outcomes via computational basis states. During inference, measurement of $R_{k,i}$ yields an outcome $j \in \{0, \ldots, b-1\}$ with probability given by
\begin{equation}
    \mathbb{P}(y_{k,i} = j) = \| \Pi_j^{(k,i)} |\psi(x; \theta)\rangle \|^2,
\end{equation}
where $\Pi_j^{(k,i)}$ is the projector associated with digit value $j$ in register $R_{k,i}$. These projectors may correspond to either disentangled or entangled subspaces, depending on the circuit architecture.

The full output vector $\tilde{y} \in \mathbb{Y}_{b,n,m}^M$ is reconstructed by performing measurements across all digit registers $\{R_{k,i}\}$ for $k = 1,\ldots,M$ and $i = -m,\ldots,n$. The resulting digits define a base-$b$ expansion of each component $\theta_k$, yielding a digitized approximation $\tilde{\theta} \in \mathbb{R}^M$ of the true latent parameter. This circuit-level digit encoding aligns naturally with the segmentation framework, where each parameter is constructed incrementally from discrete digitwise samples.

The modularity of this design enables hierarchical learning: different circuit blocks can specialize to specific digits or components, allowing reuse and parallelization. Post-training inference proceeds by repeated measurement of each register to generate empirical digit distributions, which are then refined via classical segmentation. This approach avoids the need to backpropagate through the forward model, enabling compatibility with black-box physical solvers.

From a training perspective, the circuit supports gradient-based optimization via variational methods such as the parameter-shift rule. Through repeated measurement and post-processing, the QNN can be trained to generate digit distributions aligned with structured inverse tasks~\cite{mitarai2018parameter}.

Because quantum outputs are probabilistic, digitwise errors may arise due to sampling noise, decoherence, or hardware crosstalk. Let $\theta_k$ denote the reconstructed parameter and $\theta_k^*$ the true target. The expected squared error satisfies
\begin{equation}
\mathbb{E}\left[(\theta_k - \theta_k^*)^2\right] = \sum_{i=-m}^{n} \mathbb{E}[\delta_{k,i}^2] b^{2i} + \sum_{\substack{i \neq j}} \operatorname{Cov}(\delta_{k,i}, \delta_{k,j}) b^{i+j},
\end{equation}
where $\delta_{k,i} = y_{k,i} - y^*_{k,i}$ is the digitwise error. This decomposition highlights the importance of high-precision sampling in the most significant digits, where noise has the largest effect on reconstruction accuracy. Techniques such as repeated sampling, majority voting, or entropy-aware fallback can be used to suppress high-impact digit noise and stabilize inference.
\subsection{Digit Construction via Quantum Measurement}

In the segmentation framework, each digit $y_{k,i}$ is estimated by repeatedly measuring its associated quantum register $R_{k,i}$. Let $R$ denote the number of repeated measurements. The resulting digit values follow a quantum probability distribution determined by the Born rule. From these samples, an empirical frequency distribution is computed and used to guide classical refinement.

\begin{definition}[Empirical Digit Distribution]
Let $R_{k,i}$ denote the quantum register corresponding to digit $y_{k,i}$. For $R$ repeated measurements, define the empirical distribution
\begin{equation}
    f^{(j)}_{k,i} := \frac{1}{R} \sum_{r = 1}^{R} \mathbb{I}\left[ \text{measurement} = j \right], \ \  j \in \{0, \ldots, b - 1\},
\end{equation}
where $\mathbb{I}[\cdot]$ is the indicator function. This provides a sample-based approximation to the Born probability
\[
P_{k,i}(j) = |\langle j | \psi_{k,i} \rangle|^2,
\]
for the local quantum state $|\psi_{k,i}\rangle$ of register $R_{k,i}$.
\end{definition}

Candidate digit sets $C_{k,i}$ are constructed from these empirical distributions, typically by thresholding or selecting the top-ranked values under $f^{(j)}_{k,i}$. These candidate sets form the input to the classical segmentation--elimination algorithm, which performs digitwise selection via forward model evaluation.

\begin{assumption}[Digitwise Quantum Noise Model]
\label{assumption:digit_noise}
Let \( \delta_{k,i} := y_{k,i} - y^*_{k,i} \) denote the error in the digit \( y_{k,i} \) at position \( (k,i) \), obtained from repeated quantum measurement of register \( R_{k,i} \). We assume that:
\begin{enumerate}
    \item Each \( \delta_{k,i} \) is a discrete random variable supported on \( \{-b+1, \dots, b-1\} \).
    \item The distribution of \( \delta_{k,i} \) is determined by the empirical frequency vector \( \{ f^{(j)}_{k,i} \} \) from repeated measurements.
    \item Digit errors are statistically independent across registers:
    \[
    \mathbb{E}[\delta_{k,i} \, \delta_{k',j}] = \mathbb{E}[\delta_{k,i}] \, \mathbb{E}[\delta_{k',j}]
    \quad \text{for all } (k,i) \neq (k',j).
    \]
\end{enumerate}
\end{assumption}

To analyze the impact of digitwise noise, we now formalize how uncertainty in individual digit values propagates through the reconstruction.

\begin{theorem}[Expected Squared Error under Correlated Digitwise Noise]\label{thm:noise}
Let \( \theta_k^* \in \mathbb{R} \) be the true parameter and \( \theta_k \in \mathbb{Y}_{b,n,m} \) its reconstruction via digitwise quantum measurement:
\begin{equation}
    \theta_k = \sum_{i=-m}^{n} y_{k,i} b^i, \qquad \theta_k^* = \sum_{i=-m}^{n} y^*_{k,i} b^i,
\end{equation}
with digit errors \( \delta_{k,i} := y_{k,i} - y^*_{k,i} \). Then, under Assumption~\ref{assumption:digit_noise}, the expected squared reconstruction error satisfies
\begin{equation}
    \mathbb{E}\left[ (\theta_k - \theta_k^*)^2 \right] = 
    \sum_{i=-m}^{n} \mathbb{E}[\delta_{k,i}^2] b^{2i}
    + \sum_{\substack{i,j=-m \\ i \neq j}}^{n} \text{Cov}(\delta_{k,i}, \delta_{k,j}) \, b^{i+j}.
\end{equation}
\end{theorem}

\begin{proof}
Expanding the digitwise reconstruction, we write
\[
\theta_k - \theta_k^* = \sum_{i=-m}^{n} \delta_{k,i} b^i.
\]
Then,
\[
(\theta_k - \theta_k^*)^2 = \left( \sum_{i=-m}^{n} \delta_{k,i} b^i \right)^2 
= \sum_{i,j=-m}^{n} \delta_{k,i} \delta_{k,j} b^{i+j}.
\]
Taking expectation and using linearity,
\[
\mathbb{E}\left[ (\theta_k - \theta_k^*)^2 \right] = \sum_{i,j=-m}^{n} \mathbb{E}[\delta_{k,i} \delta_{k,j}] b^{i+j}.
\]
Split into diagonal and off-diagonal terms:
\[
= \sum_{i=-m}^{n} \mathbb{E}[\delta_{k,i}^2] b^{2i}
+ \sum_{\substack{i,j=-m \\ i \neq j}}^{n} \mathbb{E}[\delta_{k,i} \delta_{k,j}] b^{i+j}.
\]
Now write
\[
\mathbb{E}[\delta_{k,i} \delta_{k,j}] = \text{Cov}(\delta_{k,i}, \delta_{k,j}) + \mathbb{E}[\delta_{k,i}]\mathbb{E}[\delta_{k,j}].
\]
If the digit errors are zero-mean, the second term vanishes, yielding the desired result.
\end{proof}

This result quantifies how digitwise noise propagates through the base-$b$ expansion. Variance in high-order digits contributes more strongly due to exponential scaling with $b^i$. Digit correlation further amplifies error through cross terms. 

In practice, digit errors may not be independent due to hardware noise, entanglement, or crosstalk. While the result above assumes unbiased, uncorrelated noise, we can extend the bound to include correlated digit errors.

\begin{corollary}
Under Assumption~\ref{assumption:digit_noise}, suppose further that if digit errors satisfy \( \mathbb{E}[\delta_{k,i}^2] \leq \sigma^2 \) and \( |\text{Cov}(\delta_{k,i}, \delta_{k,j})| \leq \rho \) for all \( i \neq j \), then
\begin{equation}
    \mathbb{E}\left[ (\theta_k - \theta_k^*)^2 \right] 
    \leq \sigma^2 \sum_{i=-m}^{n} b^{2i} + \rho \sum_{\substack{i,j=-m \\ i \neq j}}^{n} b^{i+j}.
\end{equation}
\end{corollary}

\begin{proof}
Apply the bounds directly:
\begin{align}
\sum_{i=-m}^{n} \mathbb{E}[\delta_{k,i}^2] b^{2i} &\leq \sigma^2 \sum_{i=-m}^{n} b^{2i}, \label{eq:variance_bound} \\
\sum_{\substack{i,j=-m \\ i \neq j}}^{n} \text{Cov}(\delta_{k,i}, \delta_{k,j}) b^{i+j} 
&\leq \rho \sum_{\substack{i,j=-m \\ i \neq j}}^{n} b^{i+j}. \label{eq:covariance_bound}
\end{align}
\end{proof}

This bound demonstrates that reconstruction error remains controlled under bounded variance and covariance assumptions. Moreover, repeated measurements and majority vote strategies can reduce digit error probability exponentially, even in the presence of sampling noise. For a target digitwise error rate $\epsilon$, the number of measurements required satisfies. These bounds rely on the independence and bounded-error structure specified in Assumption~\ref{assumption:digit_noise}.

\[
N \gtrsim \frac{1}{2\delta^2} \log\left( \frac{b - 1}{\epsilon} \right),
\]
where $\delta := p^* - \max_{j \neq y^*_{k,i}} P(y_{k,i} = j)$ is the confidence margin of the correct digit. This ensures the correct digit is selected with probability at least $1 - \epsilon$ under majority voting.

Digitwise segmentation naturally aligns with the modular structure of fault-tolerant quantum computing. Each register $R_{k,i}$ can be mapped to a logical qubit block, and repeated sampling serves as a shallow application-level decoding layer. Because digit errors are localized in the expansion, reconstruction is inherently robust to isolated faults—especially in lower-order digits where error contribution decays exponentially. This stands in contrast to analog quantum encodings, where errors may propagate globally.

Altogether, the segmentation--elimination framework offers a form of \emph{natural fault tolerance}, where precision is accumulated hierarchically and error is spatially confined. The digitized output model is scalable, parallelizable, and compatible with realistic NISQ-era quantum hardware, requiring only $\lceil \log_2 b \rceil$ qubits per digit and enabling robust inference through classical post-processing.
\subsection{Hybrid Quantum--Classical Segmentation Loop}

The segmentation--elimination framework operates within a hybrid quantum--classical loop that couples quantum digit sampling with classical error-based refinement. Given an input $x \in \mathbb{R}^L$, a trained quantum neural network prepares a quantum state \( |\psi(x; \theta)\rangle = U(\theta)|0\rangle^{\otimes n} \), and performs repeated measurements on each digit register \( R_{k,i} \). These measurements yield empirical distributions \( \{f^{(j)}_{k,i}\} \) over candidate digit values \( j \in \{0, \ldots, b-1\} \), approximating the Born probabilities associated with each register. From these distributions, candidate sets \( C_{k,i} \) are constructed—typically via thresholding or top-$r$ selection—to capture high-confidence quantum predictions.

Once candidate digits are generated, the classical forward model \( F(y) \) is used to evaluate each substitution \( y^{(j)}_{k,i} \) in terms of the regularized loss \( \mathcal{L}(y) = E(y) + \lambda R(y) \). The digit that minimizes this loss is selected, and the update is applied. This process is repeated across all digit positions, proceeding from most to least significant positions in a coarse-to-fine schedule.

\begin{definition}[Hybrid Segmentation Architecture]
Let \( F: \mathbb{R}^M \to \mathbb{R}^L \) be a classical forward model and \( U(\theta) \) a variational quantum circuit encoding base-$b$ digits into quantum registers \( R_{k,i} \). The hybrid segmentation loop proceeds as follows:

\textbf{(1) Quantum Sampling.} For input \( x \in \mathbb{R}^L \), prepare \( |\psi(x; \theta)\rangle \) and perform $R$ measurements of each register \( R_{k,i} \) to obtain empirical frequencies \( f^{(j)}_{k,i} \).

\textbf{(2) Candidate Extraction.} Construct $C_{k,i} \subseteq \{0, \ldots, b - 1\}$ via:
\begin{itemize}
    \item \emph{Top-$r$ selection:} retain the $r$ most frequent digits.
    \item \emph{Thresholding:} retain digits satisfying \( f^{(j)}_{k,i} \geq \tau \) for threshold $\tau$.
\end{itemize}

\textbf{(3) Classical Evaluation.} For each \( j \in C_{k,i} \), compute the perturbed output \( y^{(j)}_{k,i} = \delta^{(j)}_{k,i}(y) \), and evaluate its error:
\[
    E^{(j)}_{k,i} := \| F(y^{(j)}_{k,i}) - x_{\text{obs}} \|^2.
\]

\textbf{(4) Digit Selection.} Choose the minimizing candidate:
\[
    y_{k,i} := \arg\min_{j \in C_{k,i}} E^{(j)}_{k,i}.
\]

\textbf{(5) Update.} Replace $y_{k,i}$ with the selected value and repeat for all digits.
\end{definition}

\begin{figure}[ht]
\centering
\includegraphics[width=0.3\textwidth]{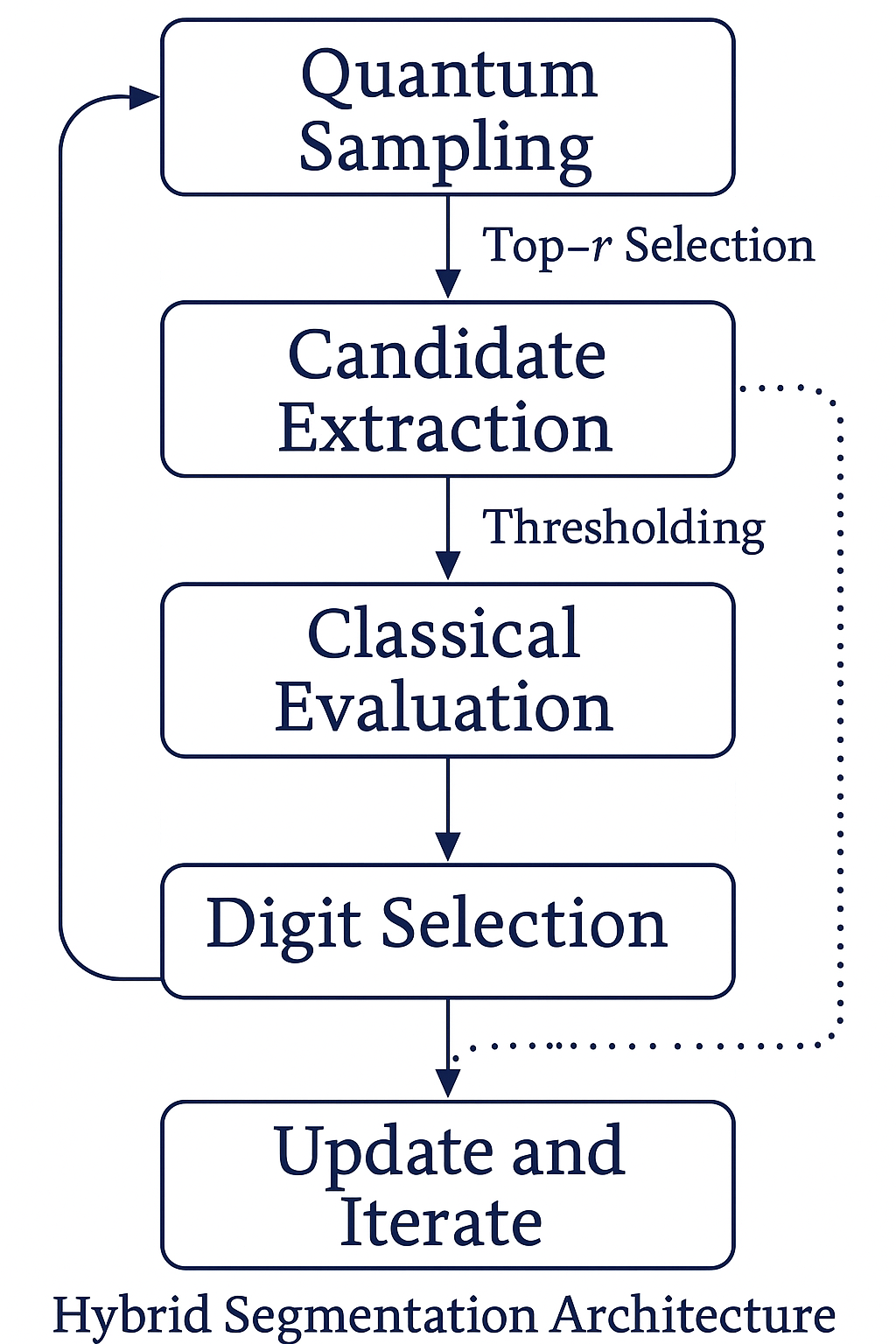} \hspace{1cm}
\caption{Flowchart of the hybrid segmentation architecture. Quantum circuits generate candidate digit values via repeated measurement. A classical forward model evaluates each candidate’s reconstruction error, and the best digit is selected. This process is iterated across digit positions, producing a high-precision hybrid estimate.}
\label{fig:hybrid-loop-flowchart}
\end{figure}

This loop defines a modular quantum--classical inference procedure. The quantum model supplies a diverse set of candidate digits via post-training sampling, while the classical model imposes physical or task-specific constraints through deterministic evaluation. Crucially, no gradients or backpropagation through the forward model are required. The architecture thus supports efficient inference in noisy intermediate-scale quantum (NISQ) settings and enables integration with black-box simulators, legacy solvers, or interpretable models. This division of roles---probabilistic generation followed by classical validation---mirrors broader trends in hybrid computing and establishes a scalable foundation for structured quantum regression.
\subsection{Resource Considerations and Parallelization}
The segmentation framework is designed for efficient execution on hybrid quantum--classical architectures, with structural features that support modular deployment, parallelization, and resource reuse. A central advantage is that quantum sampling is decoupled from classical evaluation: once the quantum circuit has been measured \( R \) times to produce empirical digit distributions \( f^{(j)}_{k,i} \), these statistics can be cached and reused throughout the digit selection phase. This avoids the need to rerun the circuit for each candidate, significantly reducing quantum resource consumption during inference.

Moreover, the quantum circuit architecture is inherently shallow. Since digits are represented as discrete classical values—rather than encoded in amplitudes or phases—the circuit depth can remain low, with minimal entanglement or decoherence. This makes the approach especially compatible with NISQ-era devices, where gate fidelity and coherence time limit circuit complexity. Shallow, register-based output also facilitates easier readout and error mitigation via repeated measurement.

The classical digit evaluation phase further supports scalability. Each candidate digit update induces an independent forward model query \( F(y^{(j)}_{k,i}) \), enabling parallel execution across processors, GPUs, or compute nodes. The digitwise structure of the parameter space allows natural partitioning, with minimal synchronization or interdependency. This enables high-throughput refinement at scale, even in large models with fine digit resolution.

Finally, the architecture’s modularity supports heterogeneous deployment. The quantum sampling module and the classical refinement logic are functionally and temporally decoupled, allowing each to run on dedicated hardware. Quantum measurements can be performed once—on dedicated quantum processors or via cloud services—and the resulting distributions streamed to classical inference engines operating asynchronously. In shared-resource environments, this separation allows quantum workloads to be batched and amortized across multiple inference tasks.

Together, these features make the segmentation--elimination algorithm resource-efficient, parallelizable, and robust to hardware limitations. By leveraging both classical and quantum strengths in a modular pipeline, the framework enables scalable hybrid inference even under the practical constraints of near-term quantum technologies.

\paragraph{Latency and Batch Inference Considerations.}
In real-time or large-scale deployment settings, latency becomes a critical concern. The segmentation framework supports batch-inference scheduling, where digit sampling and refinement stages are pipelined across parameter components or even across inference tasks. Quantum sampling for multiple inputs can be executed in parallel or sequentially batched, with empirical distributions cached and asynchronously streamed to classical refinement engines. This amortizes circuit execution costs across batches and enables efficient utilization of quantum resources. Moreover, digit layers can be updated concurrently across different parameters, supporting fine-grained pipeline parallelism that significantly reduces end-to-end latency in high-throughput inference pipelines.

\subsection{Extensions and Adaptive Strategies}
The segmentation--elimination framework supports a range of extensions that enhance its flexibility, robustness, and adaptability across computational regimes. A key generalization is \emph{adaptive base encoding}, in which the numerical base \( b \) varies across digit positions. High-order digits (larger \( i \)) may be encoded with smaller bases to reduce register width and quantum resource usage, while low-order digits (smaller or negative \( i \)) may use larger bases to improve precision. This allows fine-grained control over the trade-off between qubit overhead and reconstruction fidelity.

Another extension involves \emph{entropy-aware refinement}. When the empirical digit distribution \( f^{(j)}_{k,i} \) exhibits high entropy—indicating uncertainty or ambiguity in quantum measurements—the algorithm may trigger additional sampling to reduce variance. Alternatively, fallback heuristics such as entropy-weighted aggregation or robust voting among high-probability candidates can be applied. These mechanisms help preserve the reliability of digit selection even under noisy or flat sampling profiles.

We now formalize mixed-base digit configurations and analyze their quantization behavior.

\begin{definition}[Adaptive Digit Configuration]
Let each digit position $(k,i)$ be assigned a base \( b_k(i) \in \mathbb{Z}_{\geq 2} \). The parameter space becomes
\begin{multline}
\mathcal{Y} := \Big\{ y \in \mathbb{R}^M \ \big| \ 
y_k = \sum_{i=-m}^n y_{k,i} \, b_k(i)^i, \\
y_{k,i} \in \{0,\dots,b_k(i) - 1\} \Big\}.
\end{multline}
Each digit $y_{k,i}$ is measured from a quantum register of width \( \lceil \log_2 b_k(i) \rceil \).
\end{definition}

\begin{proposition}[Quantization Error in Mixed-Base Expansions]
Let \( \theta_k = \sum_{i = -m}^{n} y_{k,i} \cdot b_k(i)^i \) be the base-$b_k(i)$ expansion of parameter \( \theta_k \in \mathbb{R} \), and let \( \hat{\theta}_k \) be an approximation obtained by rounding or truncating each digit. Then the worst-case reconstruction error satisfies
\begin{equation}
|\theta_k - \hat{\theta}_k| \leq \sum_{i = -m}^{n} \frac{b_k(i)^i}{2}.
\end{equation}
This bound is tight when all digits are accurate except the least significant one, which differs by one unit.
\end{proposition}

\begin{proof}
Define digitwise error \( \delta_{k,i} := \hat{y}_{k,i} - y_{k,i} \), so that
\[
\theta_k - \hat{\theta}_k = \sum_{i=-m}^n \delta_{k,i} \cdot b_k(i)^i,
\]
and hence
\[
|\theta_k - \hat{\theta}_k| \leq \sum_{i = -m}^n |\delta_{k,i}| \cdot b_k(i)^i.
\]
For rounding or truncation, each \( |\delta_{k,i}| \leq \frac{1}{2} \), yielding the desired bound. If only the least significant digit differs by one unit, the bound is achieved exactly.
\end{proof}

This result shows that quantization error is dominated by low-order digits due to the geometric decay of \( b_k(i)^i \) with decreasing \( i \). High-resolution reconstruction can be achieved by increasing digit count or allocating higher bases to finer positions, though this increases qubit cost. Adaptive base scheduling thus enables efficient use of quantum resources: stable parameters may use lower bases and shorter expansions, while sensitive components receive more digits or finer base precision.

In regimes with uncertain quantum outputs, entropy-aware refinement offers a principled way to detect and correct ambiguity.

\begin{definition}[Entropy-Based Refinement Policy]
Let \( f^{(j)}_{k,i} \) denote the empirical frequency of digit value \( j \) at register \( (k, i) \), based on \( R \) repeated quantum measurements. Define the Shannon entropy:
\[
H_{k,i} := -\sum_{j = 0}^{b - 1} f^{(j)}_{k,i} \log f^{(j)}_{k,i}.
\]
A refinement policy is called entropy-aware if it triggers additional sampling or fallback heuristics whenever
\[
H_{k,i} \geq \eta,
\]
for some threshold \( \eta > 0 \).
\end{definition}

The entropy \( H_{k,i} \) of a digit register is maximized when its empirical measurement distribution is uniform, yielding \( H_{\text{max}} = \log b \). To detect uncertainty in digit predictions, a practical refinement threshold can be set as \( \eta := \log b - \delta \), where \( \delta \in (0, \log b) \) defines a tunable tolerance. For example, choosing \( \delta = \log b / 4 \) triggers refinement when the most likely digit has a probability below roughly 60--70\%. This policy identifies digit layers with diffuse or ambiguous quantum statistics and initiates corrective actions such as re-measurement to reduce entropy, soft digit interpolation among candidates, or entropy-based regularization to suppress unstable selections. By adapting computational effort to the quality of quantum outputs, entropy-aware digit refinement enhances robustness in noisy environments and increases the reliability of hybrid inference pipelines--without modifying the core structure of the segmentation algorithm.

\begin{proposition}[Mean Squared Reconstruction Error with Correlated Digit Noise]\label{prop_ed}
Let each parameter \( \theta_k \in \mathbb{R} \) be represented in a mixed-base expansion:
\[
\theta_k = \sum_{i = -m}^{n} y_{k,i} \cdot b_k(i)^i, \quad 
\tilde{\theta}_k = \sum_{i = -m}^{n} \tilde{y}_{k,i} \cdot b_k(i)^i,
\]
and define digitwise errors \( \delta_{k,i} := \tilde{y}_{k,i} - y_{k,i} \). Then the expected squared error satisfies
\begin{multline}
\mathbb{E}\big[(\tilde{\theta}_k - \theta_k)^2\big] = \sum_{i=-m}^{n} \mathbb{E}[\delta_{k,i}^2] \cdot b_k(i)^{2i} \\
+ \sum_{\substack{i,j = -m \\ i \neq j}}^{n} \mathrm{Cov}(\delta_{k,i}, \delta_{k,j}) \cdot b_k(i)^i b_k(j)^j.
\end{multline}
In particular, if \( \mathbb{E}[\delta_{k,i}^2] \leq \sigma^2 \) and \( |\mathrm{Cov}(\delta_{k,i}, \delta_{k,j})| \leq \rho \) for all \( i \neq j \), then
\begin{equation}
\mathbb{E}\big[(\tilde{\theta}_k - \theta_k)^2\big] 
\leq \sigma^2 \sum_i b_k(i)^{2i} + \rho \sum_{i \neq j} b_k(i)^i b_k(j)^j.
\end{equation}
\end{proposition}

\begin{proof}
Express the reconstruction error as
\[
\tilde{\theta}_k - \theta_k = \sum_{i = -m}^{n} \delta_{k,i} \cdot b_k(i)^i.
\]
Then the squared error expands as
\[
(\tilde{\theta}_k - \theta_k)^2 
= \sum_{i} \delta_{k,i}^2 \cdot b_k(i)^{2i} 
+ \sum_{\substack{i,j \\ i \neq j}} \delta_{k,i} \delta_{k,j} \cdot b_k(i)^i b_k(j)^j.
\]
Taking expectation,
\begin{multline}
\mathbb{E}[(\tilde{\theta}_k - \theta_k)^2] 
= \sum_i \mathbb{E}[\delta_{k,i}^2] \cdot b_k(i)^{2i} 
\\
 \sum_{i \neq j} \mathbb{E}[\delta_{k,i} \delta_{k,j}] \cdot b_k(i)^i b_k(j)^j.
\end{multline}
The second term is the sum of covariances. Applying the bounds completes the proof.
\end{proof}

Quantum measurement noise can introduce positive correlations between digit errors, particularly among entangled or adjacent qubits. These correlations may amplify overall error, posing challenges for high-precision reconstructions. A simplified bound follows under independence assumptions.

\begin{corollary}[Expected Squared Error under Adaptive Configuration]\label{cor:adaptive_config}
Under the conditions of Proposition~\ref{prop_ed}, and assuming independent, bounded digit error rates \( \epsilon_{k,i} \), the expected squared error satisfies
\begin{equation}
\mathbb{E}\big[(\tilde{\theta}_k - \theta_k)^2\big] 
\leq \sum_{i=-m}^{n} \epsilon_{k,i} \cdot b_k(i)^{2i}.
\end{equation}
If \( \epsilon_{k,i} \leq \epsilon < \tfrac{1}{2} \) and \( b_k(i) = b \) for all $i$, then
\begin{equation}
\mathbb{E}\big[(\tilde{\theta}_k - \theta_k)^2\big] 
\leq \epsilon \sum_{i=-m}^{n} b^{2i}.
\end{equation}
\end{corollary}

\begin{proof}
Under the independence assumption and zero-mean conditioned noise, the cross-terms vanish. Then
\[
\mathbb{E}[(\tilde{\theta}_k - \theta_k)^2] 
= \sum_{i=-m}^{n} \mathbb{E}[\delta_{k,i}^2] \cdot b_k(i)^{2i}.
\]
Assuming \( \mathbb{E}[\delta_{k,i}^2] \leq \epsilon_{k,i} \) gives the first bound. The uniform case follows by substitution.
\end{proof}

These bounds underscore the importance of local error control in digitwise quantum inference. They also illustrate how base scheduling interacts with noise propagation: larger bases improve resolution but amplify sensitivity to digit error, while smaller bases limit expressivity but offer resilience.

In settings where quantum sampling is unavailable or costly, segmentation can be emulated using classical generative models. Normalizing flows, masked autoregressive networks, or variational autoencoders may be trained to sample digit distributions \( f^{(j)}_{k,i} \) consistent with observed data. These synthetic distributions replace quantum outputs in the hybrid loop, allowing purely classical systems to benefit from the same digitwise probabilistic inference architecture.

At the opposite extreme, the segmentation loop may be embedded entirely within a quantum system. If the forward model \( F \) corresponds to a quantum process (e.g., Hamiltonian evolution or variational eigensolvers), the error functional \( E(y) \) can be evaluated quantumly. The refinement loop then becomes a closed quantum-classical feedback system, operating fully within a digitized quantum representation of the parameter space. This extension defines a general-purpose, fault-tolerant quantum inference primitive.

The segmentation--elimination framework thus spans a wide spectrum of computational architectures: from NISQ-compatible quantum-assisted workflows to fully classical quantum-inspired emulators and end-to-end quantum-native systems. Its modularity, fault tolerance, and hierarchical structure offer a flexible foundation for precision learning in quantum machine learning and scientific computing.

\section{Theoretical Considerations and Limitations}
\label{sec:theory}
While the segmentation-elimination framework enables digitwise reconstruction of latent parameters with remarkable numerical accuracy, it also introduces new theoretical challenges. This section formalizes certain structural properties of the algorithm and addresses its fundamental limitations. In particular, we examine the geometric implications of digitwise optimization, the risk of local minima in the non-convex error landscape, and the complexity of inference under quantum measurement noise~\cite{ge2015escaping}.

\subsection{Non-Convexity and Digitwise Coherence}

The base-\( b \) digit expansion imposes a discrete combinatorial structure on parameter space, fragmenting the domain of \( \theta \in \mathbb{R}^M \) into a collection of nested lattices. Optimization proceeds by selecting digits that locally reduce the forward error, but this process lacks a convex objective and may suffer from sharp transitions in the reconstruction loss. The local behavior of the forward model around a digit boundary is often highly nonlinear, especially in the presence of oscillatory kernels or ill-posed inverse mappings.

We formalize the potential for suboptimal convergence via the following:

\begin{proposition}[Digitwise Optimization May Converge to Local Minima]
Let \( F : \mathbb{R}^M \rightarrow \mathbb{R}^L \) be a nonlinear forward model, and let \( y \in Y_{b,n,m}^M \) denote a digitized parameter vector. Even when the true latent parameter \( \theta^\ast \notin Y_{b,n,m}^M \), there exist sequences of locally optimal digit assignments that do not globally minimize the forward error \( E(y) = \|F(y) - x_{\text{obs}}\|^2 \). In particular, if \( F \) has regions of flat gradient or discontinuous response across digit transitions, then digitwise greedy selection may converge to a suboptimal attractor basin.
\end{proposition}

\begin{proof}
Let \( F: \mathbb{R}^M \rightarrow \mathbb{R}^L \) be a nonlinear forward model, and define the error functional \( E(y) := \|F(y) - x_{\text{obs}}\|_2^2 \). Suppose the true latent parameter \( \theta^\ast \in \mathbb{R}^M \) does not lie in the digitized lattice \( Y := Y_{b,n,m}^M \subset \mathbb{R}^M \), but resides within its convex hull. The segmentation-elimination algorithm seeks an approximate solution \( y \in Y \) that minimizes \( E(y) \), typically by greedily updating one digit \( y_{k,i} \) at a time, holding all others fixed.

Since \( Y \) is a discrete subset of \( \mathbb{R}^M \), composed of a Cartesian product of digit values from finite sets \( \{0, \dots, b - 1\} \), the optimization proceeds over an isolated set of parameter configurations with no continuous interpolation between them. In such a setting, local descent is defined combinatorially: for a given digit position \( (k, i) \), we consider candidate replacements \( j \in C_{k,i} \) and accept the digit update only if it decreases the error.

Now assume that \( F \) is nonlinear and exhibits regions of flat response or non-smooth behavior—e.g., where the local gradient vanishes or jumps across adjacent digit values. Then there may exist a point \( \hat{y} \in Y \) such that no single-digit update can further reduce the error:
\begin{equation}
E(\hat{y}) \leq E\left( \delta_{k,i}^{(j)}(\hat{y}) \right) \quad \text{for all } (k,i) \text{ and all } j \in C_{k,i},
\end{equation}
where \( \delta_{k,i}^{(j)}(\hat{y}) \) denotes the digit vector obtained by replacing \( \hat{y}_{k,i} \) with \( j \) while keeping all other digits fixed.

Such a point \( \hat{y} \) is a local minimum with respect to digitwise updates, but not necessarily a global minimizer of \( E \) over \( Y \). In particular, if the true parameter \( \theta^\ast \) lies near \( \hat{y} \) but outside \( Y \), or if the error landscape contains nontrivial curvature between \( \hat{y} \) and a better digit configuration, greedy optimization may become trapped in this suboptimal basin.

This failure mode illustrates a structural limitation of digitwise descent in discrete spaces: the absence of global coordination or lookahead means that the algorithm can prematurely settle in locally optimal but globally inferior configurations. For this reason, the segmentation-elimination framework incorporates strategies such as beam search, backtracking, and stochastic digit perturbation to mitigate such convergence failures and increase the likelihood of escaping suboptimal digit sequences.
\end{proof}
\begin{proposition}[Digitwise Discontinuity of Forward Error]
Let \( F: \mathbb{R}^M \to \mathbb{R}^L \) be a continuous forward map, and define the restriction \( F|_{Y^M_{b,n,m}} \) to the digit lattice \( Y := Y^M_{b,n,m} \subset \mathbb{R}^M \). Then the composition
\[
E: Y \to \mathbb{R}_{\geq 0}, \quad E(y) := \|F(y) - x\|_2^2
\]
is piecewise-defined and generally non-smooth. In particular, small changes in digit configuration can lead to large changes in \( F(y) \), especially when \( F \) is sensitive or nonlinear. The map \( E \) is not Lipschitz-continuous over \( Y \), and may exhibit discrete jumps:
\[
\exists\, y, y' \in Y : \|y - y'\|_\infty = b^{-i},\quad |E(y) - E(y')| \gg \|y - y'\|_2.
\]
\end{proposition}

\begin{proof}
Let \( Y := Y^M_{b,n,m} \subset \mathbb{R}^M \) denote the set of digitized parameter vectors under base-\(b\) expansion, and define the forward error functional
\begin{equation}
E(y) := \|F(y) - x\|_2^2,
\end{equation}
where \( F: \mathbb{R}^M \to \mathbb{R}^L \) is a continuous forward model. Although \( F \) is continuous on the ambient space \( \mathbb{R}^M \), its restriction to the discrete subset \( Y \) inherits no such continuity. Each component \( y_k \in \mathbb{R} \) is composed of a fixed-length digit expansion
\begin{equation}
y_k = \sum_{i = -m}^{n} y_{k,i} \cdot b^i, \quad \text{with } y_{k,i} \in \{0, \dots, b - 1\},
\end{equation}
yielding a minimum nonzero step size of \( b^{-m} \) and a total of \( b^d \) possible values per component, where $d$ is the digit depth.

Consider two points \( y, y' \in Y \) that differ only in a single digit position, say \( y_{k,i} \to y'_{k,i} = y_{k,i} + 1 \), leaving all other digits fixed. This induces a change of
\begin{equation}
\|y - y'\|_2 = b^i,
\end{equation}
which may be arbitrarily small when \( i \) is negative. However, if the forward model \( F \) is sensitive to perturbations in this component, then
\begin{equation}
|E(y) - E(y')| = \left| \|F(y) - x\|_2^2 - \|F(y') - x\|_2^2 \right|
\end{equation}
may be significantly larger than \( \|y - y'\|_2 \). In particular, the error landscape may exhibit large jumps across digit boundaries even as the underlying change in parameter values vanishes.

Since the set \( Y \) is discrete, no open ball around any point is fully contained within \( Y \), and the restriction \( E|_Y \) is not continuous in the topological sense. Moreover, it fails to satisfy any global Lipschitz condition, as the ratio
\begin{equation}
\frac{|E(y) - E(y')|}{\|y - y'\|_2}
\end{equation}
can diverge as \( i \to -\infty \). Therefore, although \( F \) may be smooth on \( \mathbb{R}^M \), its composition with the digitized lattice structure induces a non-smooth, piecewise error surface over \( Y \). This breakdown of continuity explains the necessity of beam search, backtracking, and stochastic digit refinement in the segmentation-elimination algorithm, which must navigate the resulting nonconvex and discontinuous geometry.
\end{proof}

\paragraph{Mitigation via Entropy-Aware Digit Refinement.}
The discontinuities and local traps described above can be mitigated by augmenting the segmentation-elimination procedure with entropy-aware refinement strategies. When a digit register exhibits a high-entropy empirical distribution---indicating uncertainty or the absence of a dominant candidate---the algorithm can trigger localized grid search or fallback mechanisms based on smoothed digit distributions. These adaptive responses allow the optimizer to escape degenerate neighborhoods and partially restore continuity within the otherwise discrete and fragmented error landscape. By monitoring digitwise entropy and local variance, the algorithm dynamically adjusts its search granularity and sampling effort, thereby improving robustness in regions of non-smooth or unstable forward behavior.

Even when the forward model \( F \) is differentiable on the ambient space \( \mathbb{R}^M \), its restriction to the discrete digit lattice \( Y \) induces a non-smooth and often discontinuous response surface. Small perturbations to digit values can lead to abrupt changes in the forward prediction \( F(y) \), especially near digit boundaries, resulting in sharp transitions or plateaus in the loss functional \( E(y) \). These effects hinder both greedy optimization and gradient-based refinement.

To overcome this, the segmentation-elimination algorithm incorporates global convergence strategies such as beam search, multi-digit backtracking, and coarse-to-fine refinement. In hybrid quantum-classical implementations, additional smoothing can be introduced through soft digit selection or interpolation over candidate neighborhoods, further improving stability.

These enhancements address the empirically observed failure modes of naive greedy segmentation, which can stall in shallow or deceptive basins of the loss landscape. As detailed in Section~\ref{sec:toy_prob}, global convergence strategies---including entropy-aware digit pruning, variable-width beam search, digit penalty regularization, multiscale refinement, and annealed digit perturbation---significantly reduce the likelihood of convergence failure. When applied in concert, they enable robust recovery of latent parameters even in the presence of nonconvexity, digit saturation, or measurement noise.
\subsection{Limits of Quantum Sampling and Noise Tolerance}
The digitwise segmentation framework fundamentally relies on the assumption that each measured digit \( y_{k,i} \) corresponds either to a deterministic outcome or to a probabilistic sample drawn from a stable, repeatable distribution. In particular, the framework presupposes that digit-level outputs are accurate within a known noise tolerance, either through high-fidelity measurement or through repeated sampling that averages out uncertainty. However, this assumption becomes fragile under realistic quantum hardware constraints. Quantum measurement is inherently stochastic, and in near-term devices, errors may be not only noisy but also biased, temporally correlated, or spatially entangled across qubit registers. These deviations from idealized error models can distort the empirical digit distributions \( f^{(j)}_{k,i} \), leading to misclassification even when a large number of samples are collected.

Section~5 provided theoretical guarantees (Theorem~\ref{thm:noise} and Corollary~\ref{cor:adaptive_config}) under the simplifying assumption of independent digitwise errors with fixed misclassification probability \( \varepsilon < \frac{1}{2} \). These results demonstrated that applying majority vote across repeated measurements suppresses digitwise error exponentially in the number of repetitions \( R \), effectively converting probabilistic readouts into stable classical values. Yet this strategy has practical limits. In the presence of coherent or systematic noise--such as readout cross-talk, gate drift, or correlated environmental decoherence--the independence assumption fails, and majority vote may merely reinforce biased outcomes~\cite{geller2020rigorous}. Moreover, for low-entropy digit registers--those with flat, uncertain distributions over many possible digit values--the number of repetitions required to reliably extract the correct digit can become prohibitively large, especially in bases \( b > 4 \) where the decision boundary between modes is narrow.

These considerations expose a fundamental triad of trade-offs in digit-level quantum inference: precision, robustness, and cost. Increasing numerical precision by raising the base \( b \) or extending the digit depth $d$ expands the effective parameter resolution, but it also increases the number of qubits needed per digit register and makes each digit more sensitive to measurement error. On the other hand, increasing robustness by performing majority vote across more samples improves reliability, but it comes at the cost of increased runtime and quantum sampling overhead. The segmentation framework accommodates flexible tuning across these axes--each digit position can be configured independently in terms of base, sampling depth, and refinement strategy--but choosing an optimal configuration requires detailed knowledge of hardware capabilities, circuit fidelity, and the sensitivity structure of the forward model \( F \).

In practice, optimal deployment may involve heterogeneous digit strategies: high-significance digits may be sampled more aggressively to ensure stability, while lower-order digits--less influential on the overall parameter--can be estimated with coarser resolution or reduced sampling. Ultimately, digitwise segmentation transforms the quantum inference process into a structured and interpretable optimization problem, but its reliability depends critically on managing and correcting the nuanced imperfections introduced by real-world quantum hardware.
\subsection{Toward Fully Quantum Segmentation and Learning}
A final speculative direction envisions embedding the segmentation-elimination framework into fully quantum end-to-end learning pipelines. In the current hybrid implementation, quantum circuits are primarily used to generate digit-level samples for latent parameters, while classical routines perform forward evaluation, digit selection, and error-driven refinement. However, if both components of the loop--the digit generation module and the forward model--can be implemented as quantum subroutines, then the entire inference and optimization cycle may be promoted to a quantum-native feedback system.

In this envisioned architecture, quantum circuits would not only generate candidate digit configurations but also simulate or approximate the forward model \( F \) within a quantum computing framework. This is feasible for certain classes of problems: for instance, when \( F \) represents the time evolution of a quantum system, the solution to a PDE via a Hamiltonian operator, or a variational energy functional in quantum chemistry, then the quantum forward map can be constructed using established techniques such as Trotterized evolution, VQE, or qEOM solvers. In such a setting, forward error evaluation would proceed via expectation value estimation or phase estimation circuits, and the segmentation loop--rather than calling a classical function \( F(y) \)--would estimate \( E(y) \) quantumly.

With the segmentation loop now residing entirely within quantum space, digit candidate evaluation becomes a quantum search problem. This opens the door to using amplitude amplification techniques (such as Grover-like oracles) to identify candidate digits with minimal error, possibly over superpositions of candidate configurations. Furthermore, if one can encode digit sequences as quantum registers and propagate updates through a variationally trained policy, one could implement quantum digit selection strategies using quantum reinforcement learning or unitary policy networks, bypassing classical digit scanning altogether. Such circuits might update digit assignments through entanglement-driven dynamics, effectively learning error-minimizing transitions in Hilbert space.

The practical realization of such an architecture requires advances in several areas. First, mid-circuit measurement and reset capabilities must become more reliable and scalable, enabling dynamic digit updates without collapsing the full quantum state~\cite{murali2020software}. Second, quantum memory or modular register preservation is needed to store partial digit reconstructions between steps. Third, quantum resource overhead (e.g., in circuit width, ancilla management, and noise mitigation) must be significantly reduced via fault-tolerant encoding and quantum compilation strategies~\cite{roberts2020memory}. Nonetheless, these challenges are already active areas of research in the quantum computing community.

If realized, a fully quantum-native segmentation-elimination framework would represent a novel intersection of discrete optimization, digit-encoded inference, and quantum feedback control. Unlike many quantum learning algorithms that rely on dense continuous encodings or abstract kernel representations, digitwise segmentation offers an interpretable, modular, and inherently hierarchical structure. It transforms the problem of quantum inference from one of function approximation into a structured combinatorial process--with each digit corresponding to a meaningful decision node in a quantum circuit. As such, it would provide a new class of models for discrete-valued learning in quantum machine learning, bridging symbolic reasoning and quantum variational computation in a highly interpretable way.

Segmentation-elimination provides a powerful and adaptable primitive for inverse recovery in hybrid quantum-classical settings. However, its success depends on careful management of digitwise uncertainty, global error geometry, and sampling complexity. While current implementations are effective for structured inverse problems and scientific computation, many foundational questions remain open--offering fertile ground for future theoretical investigation.

\section*{Final Remarks}
This work has introduced and systematically developed a segmentation-based regression framework tailored to quantum neural networks (QNNs), enabling high-precision inference over discretized output spaces. By encoding real-valued targets as finite base-$b$ digit expansions, the regression task is reformulated as a structured combinatorial optimization over a discrete digit lattice. This formulation transforms continuous parameter estimation into a tractable digitwise search problem, allowing for scalable and interpretable regression.

At the core of the method is the segmentation--elimination algorithm, which incrementally refines a candidate solution by locally minimizing a forward error functional at each digit position. This hierarchical digitwise construction aligns naturally with the discrete nature of quantum measurement and is compatible with near-term quantum hardware. Quantum circuits are used to generate digit candidates probabilistically, while classical forward models evaluate these candidates to guide selection. This hybrid architecture cleanly decouples stochastic generation from deterministic evaluation, allowing each computational component to operate within its respective strengths.

We have established theoretical guarantees on convergence, digitwise local optimality, and runtime complexity, while also analyzing the method’s limitations under nonconvexity, digit clipping, and measurement noise. Applications to scientific inverse problems, particularly those governed by partial differential equations (PDEs), demonstrate the framework’s practical viability in data-driven modeling, parameter estimation, and nondifferentiable simulation contexts. The approach supports parallelism, requires no gradient information, and interfaces naturally with legacy solvers; making it well-suited for modular deployment in scientific pipelines.

Beyond its immediate applications, segmentation-based regression opens a range of promising avenues for future research. One direction involves incorporating global search strategies; such as beam search, simulated annealing, or evolutionary heuristics, to improve convergence in complex, nonconvex landscapes and enhance robustness against local minima. Another direction focuses on uncertainty-aware refinement, where entropy or confidence estimates guide adaptive allocation of computational resources toward ambiguous or high-impact digit positions. A more advanced extension envisions embedding the segmentation framework within quantum-native forward models, including variational eigensolvers and quantum PDE solvers, thereby enabling end-to-end inference loops entirely within quantum computational architectures. Additionally, the discrete and interpretable nature of digitwise regression aligns naturally with symbolic and digital control applications, such as controller synthesis, filter design, and program generation, where structured outputs are essential.

In summary, the segmentation-elimination framework provides a principled, interpretable, and scalable approach to digit-level regression in hybrid quantum-classical settings. By bridging the representational gap between quantum measurement outputs and the continuous demands of scientific inference, it lays a rigorous foundation for quantum-enhanced learning pipelines with real-world impact.
\bibliographystyle{abbrv}
\bibliography{bibliography}

\end{document}